\documentclass[12pt]{article}
\newcommand{\be}{\begin{eqnarray}}
\newcommand {\ee}{\end{eqnarray}}
\newcommand{\aq}{\ensuremath{\mathcal{A}_{Q}}}
\newcommand{\hs}{\ensuremath{\mathcal{H}}}
\newcommand{\acl}{\ensuremath{\mathcal{A}_{cl}}}
\newcommand{\cim}{\ensuremath{C^{\infty}(M)}}
\newcommand{\sca}{\ensuremath{\mathcal{A}}}
\newcommand{\scb}{\ensuremath{\mathcal{B}}}
\newcommand{\scx}{\ensuremath{\mathcal{X}}}
\newcommand{\scy}{\ensuremath{\mathcal{Y}}}
\newcommand{\sct}{\ensuremath{\mathcal{T}}}
\newcommand{\ax}{\ensuremath{(\mathcal{A}, \mathcal{X})}}
\newcommand{\by}{\ensuremath{(\mathcal{B}, \mathcal{Y})}}

\newcommand{\axal}{\ensuremath{( \mathcal{A}, \mathcal{X}, \alpha )}}
\newcommand{\bybe}{\ensuremath{( \mathcal{B}, \mathcal{Y}, \beta )}} 
\newcommand{\sone}{\ensuremath{\mathcal{S}_{1}}}
\newcommand{\oq}{\ensuremath{\omega_{Q}}}
\begin{document}
\begin{center}
{\Large Towards an Autonomous Formalism for Quantum Mechanics}\\
\vspace{.2in}
\textbf{ Tulsi Dass} \\
\vspace{.2in}
Department of Physics, Indian Institute of Technology, Kanpur, India 208016 \\
E-mail: tulsi@iitk.ac.in
\end{center}
\vspace{.5in}

  A formalism is presented in which quantum particle dynamics can be developed on its own rather
than `quantization' of an underlying classical theory. It is proposed that the unification of 
probability and dynamics should be considered as the basic feature of quantum theory. Arguments are 
given to show that when such a unification is attempted at the configuration space level, the 
wave funtions of Schr$\ddot{o}$dinger theory appear as the natural candidates for the desired unification. 
A *-algebra  $\mathcal{A}_{Q} $ of (not necessarily bounded) linear operators acting on an appropriate 
dense set of these wave functions appears as the arena for quantum kinematics. A simple generalization of
an existing formalism in noncommutative geometry is employed to develop the notion of generalized 
algebraic symplectic structure ( GASS ) which can accomodate classical and quantum symplectic 
structures as special cases. Quantum kinematics and dynamics is developed in in the framework of 
a noncommutative Hamiltonian system employing an appropriate GASS based in $ \mathcal{A}_{Q} $. The 
Planck constant is introduced at only one place -- in the quantum symplectic form; its appearance 
at conventional places is then automatic. Unitary Wigner symmetries appear as canonical transformations 
in the noncommutative Hamiltonian system. A straightforward treatment of quantum - classical 
correspondence is given in terms of appropriate GASSes.

\newpage
\noindent
\textsl{Contents}
\begin{description}
\item[I.]\textsl{Introduction.}
\item[II.]\textsl{Preliminaries.} A. General probabilistic description of systems. B. Probabilistic 
Hamilton - Jacobi theory (the Hamilton - Jacobi fluid). 
\item[III.] \textsl{Unification of dynamics and probability in configuration space; rationale for a 
complex Hilbert space. }A. States. B. Observables. C. Transition probabilities; fundamental 
invariances. 
\item[IV.] \textsl{Generalized algebraic symplectic structures.} A. Algebras and derivations. B. Generalized algebraic differential forms. C. Induced mappings; Lie derivatives. D. Generalized algebraic symplectic structures. E. Canonical symplectic structure on a special ADS. F. Dynamical systems in algebraic setting; generalized algebraic Hamiltonian systems. 
\item[V.] \textsl{Quantum mechanics of a particle.} A. The quantum symplectic form. B. The Galilean 
group. C. Fundamental observables. D. Dynamics. E. The Hamiltonian operator. 
\item[VI.] \textsl{Quantum - classical correspondence.} 
\item[VII.] \textsl{Concluding remarks.}
\end{description}
\textsl{ Appendix. } Symplectic Manifolds.

\begin{center}
I. INTRODUCTION
\end{center}

  The traditional formalism of quantum mechanics ( QM ) has two unsatisfactory features: (i) One all the time `quantizes' classical systems. (ii) The languages employed in the traditional treatments of QM and CM (classical mechanics) are very different; this obscures the parent - daughter relationship between the two theories.

  This is in marked contrast with the situation with some other pairs of parent - daughter 
theories in physics -- for example, the pair [ special relativistic mechanics (SRM), Newtonian 
mechanics (NM) ] and the pair (general relativity, Newtonian theory of gravitation ).In these 
cases \\
(i) the concepts and equations of the parent theory can be developed on their own; \\
(ii) there exists a general formalism such that \\
(a) both the theories can be described in it; \\
(b) basic concepts and equations of the daughter theory can be obtained from those of the parent theory in a suitable limit, and \\
(c) a suitable subclass of quantities in the parent theory go over to well defined quantities in
the daughter theory ( with a corresponding interpretation )in the limit mentioned in (b).

  In the pair SRM - NM, for example, the concepts and equations of SRM can be developed quite 
independently of NM to which SRM reduces in the $ c \rightarrow \infty $ limit. In this limit the 
relativity of simultaneity reduces to absolute simultaneity, the proper time differential $ d\tau $ 
goes over to the Newtonian time differential dt, the spatial components of the equation  
$ dp^{\mu}/d\tau = K^{\mu} $ goes over to Newton's second law, the particle energy $ E = m_{0}c^{2}/\sqrt{1-v^{2}/c^{2}} $ goes to infinity whereas the kinetic energy $ K = E - m_{0}c^{2} $ goes over to the nonrelativistic kinetic energy $ \frac{1}{2}m_{0}v^{2} $ etc.  

  The main objective of this paper is to remove the above - mentioned deficiencies in the treatment 
of QM and present a formalism in which one has an autonomous development of QM which permits a 
transparent treatment of quantum - classical correspondence.In this work we shall mainly consider 
particle QM.  
A similar autonomous treatment of quantum field dynamics  is being 
postponed to a future publication.
 
  To see the quantum to classical transitionin a transparent form, one must translate the 
operator - theoretical formalism into an `equivalent' one involving phase space functions 
(replacing commutators by the so - called Moyal brackets ). One of the main concerns of the present 
work will be to ensure that the formalism evolved permits the description of the `equivalence' 
referred to above as an isomorphism betweem appropriate mathematical structures.

  What is required is a mathematical formalism employing objects of a sufficiently general type so
as to include, as special cases, the algebra of phase space functions with the Poisson bracket 
structure defined on it and an  algebra of operators in the quantum mechanical Hilbert space with 
the Heisenberg commutators as analogues of Poisson brackets. Noncommutative geometry (NCG) [1--4] 
(in which all geometry is developed in the framework of algebras ) holds the key to the evolution 
of such a formalism.  Indeed, noncommutativity is the hallmark of QM. In the paper that marked the 
birth of QM [5], Heisenberg's main conlusion, based on correspondence arguments, was that the 
kinematics underlying QM must be based on a noncommutative algebra of observables. An intuitive 
formulation of noncommutative Hamiltonian mechanics ( matrix mechanics ) was given in ref [6-8]. 
Noncommutative symplectic structures  yielding quantum mechanical commutators as Poisson brackets 
have appeared in literature[2--4,9];  however, a mathematically satisfactory treatment of quantum 
symplectics meeting the above mentioned needs is yet to be given. Pursuit of this objective led the 
present author to a generalization [10] of the NCG scheme of ref [2,3] which holds promise for 
interesting applications. Here we shall present this generalized formalism for NCG, use it to
 evolve a general framework for mechanics ( which covers CM and QM as special cases ) and give a 
straightforward treatment of quantum - classical correspondence in  this framework.  
 
  The conceptual development of a fundamentally new theory often takes place around a unifying 
principle. For example, Maxwell-Lorentz electrodynamics unifies electricity and magnetism, special 
relativity unifies the concepts of space and time and general relativity unifies space-time 
geometry and gravitation. Is there a unifying principle underlying quantum mechanics ? One might 
suggest unification of wave and particle properties of matter. ( Indeed, this was the theme 
underlying the work of de Broglie [11] and Schr$\ddot{o}$dinger [12]. ) There is, however, in the 
author's opinion, a deeper unification -- that of dynamics and probability -- which, when 
incorporated in an appropriate framework, entails the unification of wave and particle properties 
of matter. One possible way to understand it is this: There is a background noise field pervading 
all universe whose dynamics is inexorably tied up with that of matter such that the effect of of 
this noise field on matter cannot be adequately treated as a perturbation. This effect 
is presumably best treated by employing, in the description of dynamics of matter, mathematical 
entities that give a unified description of dynamics and probability. The resulting description of 
material objects is expeted to involve a blend of particle - like motion with wave-like fluctuations. 

  In everyday use of QM, such a unification of probability and dynamics is taken for granted; 
indeed, we all the time employ the Schr$\ddot{o}$dinger wave functions for statistical averaging 
as well as for describing the dynamics of atomic systems. A point brought out 
in the present work (section III) is that, when such a unification is attempted at the configuration space level in an appropriate setting, Schr$\ddot{o}$dinger type wave functions appear as the natural candidates for such a unification, thus providing a rationale for the use of complex Hilbert spaces in QM.

  Earlier works relating to the foundations of QM have been generally concerned with 
quantum logic [13--15], $C^{*}$-algebras [16--21] or stochastic mechanics [22--25]. These works 
have provided valuable insights into various aspects of QM; however, as far as the question of 
providing the initial motivation for the Schr$\ddot{o}$dinger wave function is concerned, as we 
shall see in section III, the approach adopted here is much more direct and intuitively appealing. 
The formalism developed in section V is essentially algebraic; however, we do not restrict 
ourselves to $C^{*}$-algebras.

  The plan of the remaining sections is as follows. Section II is devoted to some preliminaries 
relating to the general probabilistic description of dynamics of systems [26, 27] and the 
probalistic version of Hamilton-Jacobi theory (called Hamilton-Jacobi fluid in ref [25, 28] ).In 
section III, we consider unification of dynamics and probability in particle dynamics in 
configuration space and  show that the appropriate single mathematical object unifying the 
Hamilton-Jacobi function S(x,t) and the probability density function $\rho(x,t) $  must be a 
Schr$\ddot{o}$dinger type wave function $\psi$(x,t) which is `essentially' 
$\sqrt{\rho}exp[iS/\hbar]$; the  probability interpretation of $\psi$ is then automatic. The 
physically realizable wave functions 
belong to a dense subset $\Omega$ of the Hilbert space $\mathcal{H} = L^{2}(R^{3})$ consisting of 
an appropriate class of smooth functions. A *-algebra $\mathcal{A}_{Q}$ of (not necessarily 
bounded ) operators mapping $\Omega$ into itself appears as the natural arena for quantum 
kinematics. After presenting the detailed treatment of GASSes as mentioned above (with 
$\mathcal{A}_{Q}$ as the underlying algebra )in section IV, we develop quantum kinematics and 
dynamics of a particle in the framework of a GASS based in $\mathcal{A}_{Q}$ in section V.  Quantum-classical 
correspondence is treated in sectin VI. The last section contains some concluding remarks.   

  The paper has been written in an easy style with the hope that students just being introduced 
to QM will enjoy reading a substantial part of it. We shall generally maintain a pretension that 
we are at the pre-1925 stage when the papers of Heisenberg and Schr$\ddot{o}$dinger had not 
appeared and  evolve an autonomous formalism for QM from scratch.

  A preliminary version of the essential ingredients in section IV and VI appeared in the article 
[10] which was given a limited circulation back in 1993. A brief acount of the same (along with 
some background material on classical symplectic geometry) also appeared in ref [29].

\begin{center}
II. PRELIMINARIES
\end{center}

  In this section we shall collect some useful results relating to probabilistic description of
systems and the probilistic version of Hamilton-Jacobi theory.

\noindent
A. \textsl{General Probalistic Description of Systems}

  A substantial part of this subsection is an adaptation from ref [26, 27].

  In every physical theory, there are three primitive elements [30,31] which are always present
(explicitly or implicitly ):\\
(i) observations/measurements; \\
(ii) description of evolution of systems (typically in terms of a discrete or continuous parameter called time ); \\
(iii) conditional predictions about systems : given some information about a system ( typically 
in terms of values of appropriate measurable quantities), to make predictions/retrodictions about its behaviour.

  A physical theory generally involves additional ingredients in the form of hypotheses 
concerning the relevant aspects of nature; the structure of the theory is then developed into a consistent mathematical formalism based on those hypotheses and incorporating the above mentiond three elements.

  A useful concept which serves to integrate the three items above is that of a \textit{state}.
A state of a system encodes, in a conveniently usable mathematical form, the  information 
available about the system at a particular time. A standard way to describe the evolution of a 
system is to describe the change of state with time. At the initial stages in its formulation, the main business of any 
physical theory is to provide appropriate mathematical description of observable/measurable 
quantities ( generally called observables ) and states and the relationship(s) between them.

    We shall be mainly concerned with the nonrelativistic kinematics and dynamics of a particle. In 
Newtonian mechanics, a particle is characterized by a Galilean invariant parameter m called its 
mass [32, 33]. Its motion in space is describd by giving its position vector $\vec{r}(t)$ as a 
function of time which is assumed to be smooth so that one can talk about velocity 
($\vec{v} = d\vec{r}/dt$), 
acceleration ($\vec{a} = \frac{d^{2}\vec{r}}{dt^{2}} $) etc. The equation of motion (Newton's second law ) 
gives acceleration in terms of force which is supposedly a function of position and 
velocity. We can, therefore, take, at each time t, the vectors $\vec{r}(t)$ and $\vec{v}(t)$
[or, equivalently, $\vec{r}(t)$ and $\vec{p}(t) = m \vec{v}(t)$] as independent observables. These 
are 
the fundamental observables; all other observables ( angular momentum, kinetic energy etc ) are 
functions of these. The state of the particle at time t is given by the pair ($ \vec{r}(t), 
\vec{p}(t)$) and its change with time is given by the equations
\begin{eqnarray}
\frac{d\vec{r}(t)}{dt} = \frac{1}{m}\vec{p}(t) \hspace{.3in} \frac{d\vec{p}(t)}{dt} = \vec{F}(\vec{r}(t), \vec{p}(t), t) .
\end{eqnarray}
Given the initial state ($\vec{r}(0), \vec{p}(0) $), at any later time t, the state ($\vec{r}(t), 
\vec{p}(t)$) ( and, therefore, the values of all other observables at time t) can be found by 
integrating eqs (1): we have a deterministic theory.

  The mechanics underlying atomic phenomena cannot be deterministic. Given, for example, an atomic 
electron in a sufficiently high excited Bohr energy state, it can generally make transition to more 
than one possible lower energy states; a theory of atomic phenomena is expected to predict 
probabilities of these transitions. Moreover, if radiation is assumed to consist of photons, a 
description of interference and diffraction of light can be given only in a probalistic framework 
[34].

  A probalistic situation can arise in a deterministic theory if the initial state is inadequately 
specified. This is the case in classical statistical mechanics where a state is generally represented 
by a probability density $\rho$(q, p, t) in phase space. The single particle states ( $\vec{r}(t), 
\vec{p}(t) $) mentioned above are special cases of this corresponding to
\begin{eqnarray}
\rho(\vec{r}, \vec{p}, t) = \delta( \vec{r} - \vec{r}(t) ) \delta(\vec{p} - \vec{p}(t) ) .
\end{eqnarray}
A formalism of this type, however, is not adequate [34] for the treatment of quantum phenomena.To the 
best of understanding achieved to to - date, the dynamics of atomic and subatomic systems is 
intrinsically/irreducibly probabilistic.

  What is the appropriate framework for the treatment of such dynamics ? We shall obtain below a 
standardised form for the description of states and observables in a probabilistic framework which 
is sufficiently general for our purpose.

  Consider a typical experiment in which a system is prepared with some prescribed initial 
conditions and a measurement of some quantity is made. The measured values are generally 
expected to be measurable subsets in  a measurable space ( S, $\mathcal{B}$(S)) where S is a subset of some 
Euclidean space and  $\mathcal{B}$(S) is its Borel $\sigma$-algebra.  For example, measured values of a length (with 
error margins ) are Borel subsets of the real line; measured values of a direction are Borel 
subsets of the unit sphere etc. In the latter case, the direction measurements can be 
analysed in terms of two real parameters $ \theta $ and $ \phi $. More generally, all 
measurements can be analysed in terms of simpler measurements  giving single real numbers 
( with appropriate error margins ) as measured values. For the treatment of fundamentals, it is generally adequate to consider only these simpler measurements. We shall, therefore assume 
henceforth that S is a subset of the real line.   

   We shall start by trying to give an operational meaning to the probability 
$w^{Q}_{\alpha}$(B) of an observable Q having values in a Borel set B when the system is in 
a state  $\alpha$.

  Let the experiment be performed with an experimental setup or instrumnt $\tilde{Q}$ with the 
system prepared so as to satisfy some initial conditions $\tilde{\alpha}$. In each performance of 
the measurement, the measured value will lie in some Borel subset of S. Let the experiment be repeated 
a large number (N) of times. If the measured value lies in a Borel set B in $N_{B}$ runs of the 
experiment, we have, adopting the traditional relative frequency definition of probability, an 
operationally defined probability
\begin{eqnarray}
w^{\tilde{Q}}_{\tilde{\alpha}} (B) \sim \frac{N_{B}}{N} \hspace{.2in} ( N\  large ) .
\end{eqnarray}

  We now define an equivalence relation $ \approx $ on the set $ \Sigma $ of all initial 
conditions : $ \tilde{\alpha}_{1} \approx \tilde{\alpha}_{2} $ if $ w^{\tilde{Q}}_{\tilde{\alpha}_{1}} (B) = w^{\tilde{Q}}_{\tilde{\alpha}_{2}} (B) $ for all $ \tilde{Q}$ and B. We 
denote the equivalence class of $ \tilde{\alpha}$ as $ \alpha$ and the set of equivalence 
classes in $ \Sigma $ as $ \mathcal{S} $; the members $ \alpha, \beta, ... $of 
$ \mathcal{S} $ will be called states. We have, at this stage, the quantities $ w^{\tilde{Q}}_{\alpha} $(B) defined.

  We next define an equivalence relation $ \equiv $ on the set $ \mathcal{I} $ of all 
instruments : $ \tilde{Q}_{1} \equiv \tilde{Q}_{2} $ if $ w^{\tilde{Q_{1}}}_{\alpha}(B) = 
w^{\tilde{Q_{2}}}_{\alpha}(B) $ for all $ \alpha $ and B. We denote the equivalence class of 
$ \tilde{Q} $ as Q and the set of equivalence clases in $ \mathcal{I} $ as $ \mathcal{O} $; 
the members P, Q, ... of $ \mathcal{O} $ will be called observables. A quantity
$w^{Q}_{\alpha}$(B) can now be defined; this is  the desired probability.

  Given Q and $\alpha$, we have the probability measure $w^{Q}_{\alpha}$ defined on the measurable 
space $(S, \mathcal{B}(S)$ thus obtaining a probability space $(S, \mathcal{B}(S), w^{Q}_{\alpha})$.
 The expectation value of the observable Q in the state $\alpha$ is defined as
\begin{eqnarray}
\alpha(Q) = \int_{S}sdw^{Q}_{\alpha}(s) .
\end{eqnarray}

  Given an observable Q and a Borel - measurable function f on the real line, we can define 
the function f(Q) of Q as the observable which takes a value f(q) whenever Q takes a value q.
The expectation value of f(Q) in the state $ \alpha $ is given by
\begin{eqnarray}
\alpha ( f(Q) ) = \int_{S} f(s) dw^{Q}_{\alpha} (s) .
\end{eqnarray}

  Given the expectation values $ \alpha (Q) $, we can find the quantities $ w^{Q}_{\alpha}(B)$ 
by the rule
\begin{eqnarray}
w^{Q}_{\alpha} (B) = \alpha (\chi_{B}(Q)) 
\end{eqnarray}
where $ \chi_{B} $ is the characteristic function of the Borel subset B of S.

 We next consider the structure of a `mixture space' [26] on the state space $\mathcal{S}$. Given a 
finite set \{ $\alpha_{i}$\} and weights \{$p_{i}$\} ($ 0 \leq p_{i} \leq 1, \sum_{i}p_{i} = 1$),
 the joint collection ( \{$\alpha_{i}$\}, \{$p_{i}$\}) defines a state $ \alpha$ by the relation
\begin{eqnarray*}
w^{Q}_{\alpha}(B) = \sum_{i}p_{i} w^{Q}_{\alpha_{i}}(B) \hspace{.2in} for\  all\  Q, B.   (6A)
\end{eqnarray*}
The state $\alpha$ is interpreted as a mixed state in which the state $\alpha_{i}$ appears with a 
probability $p_{i}$. The expectation value of an observable Q in this mixed state is given by
\begin{eqnarray*}
\alpha(Q) = \sum_{i}p_{i}\int_{S}sdw^{Q}_{\alpha_{i}}(s) .
\end{eqnarray*}

  A general mixture space is defined as a set M with the property that any finite collection 
\{$\mu_{i}$\} of elements of M and weights $ \{p_{i}\} $ defines a unique element $\mu$ of M subject 
to the condition that if $ \mu_{i} = \mu_{0}$ for all i, then $ \mu = \mu_{0}$. The set 
$\mathcal{S}$ is clearly a mixture space. [$\alpha_{i} = \alpha_{0} $ for all i in eq(6A) implies 
$\alpha = \alpha_{0}$.]

  An important example of a mixture space is a convex set K of a real vector space V. [ Given 
points (i. e. vectors )$K_{i}$ lying in K, we have $\sum_{i}p_{i}K_{i}$ lying in K.] We shall call such a mixture 
space a standard mixture space.

  We shall now show [26] that the space $\mathcal{S}$ can be taken, without loss of generality,
to be a standard mixture space. We shall do this by showing a one-to-one correspondence between
$\mathcal{S}$ and a standard mixture space which preserves mixtures.

  An affine functional on $\mathcal{S}$ is a mapping $\phi$ from $\mathcal{S}$ to the real line
R such that, for a mixture $ \alpha = ( \{\alpha_{i}\}, \{p_{i}\}) $ we have 
\begin{eqnarray}
\phi (\alpha ) = \sum_{i}p_{i} \phi( \alpha_{i}) .
\end{eqnarray}
The set $\mathcal{F}$ of affine functionals on $\mathcal{S}$ is easily seen to be a real vector 
space. Let $\mathcal{F}^{*}$ be the algebraic dual of $\mathcal{F}$, i.e. the space of linear 
functionals on $\mathcal{F}$; this space is also a real vector space. A one-to-one correspondence 
of $\mathcal{S}$ onto a convex subset of $\mathcal{F}^{*}$ is obtained by associating with a state 
$\alpha$ , the element $\hat{\alpha}$ of $ \mathcal{F}^{*}$ given by
\begin{eqnarray}
\hat{\alpha} (\phi) = \phi ( \alpha ) \hspace{.2in} for\  all\  \phi\  in\  \mathcal{F}.
\end{eqnarray}     
The corresponence $ \alpha \leftrightarrow \hat{\alpha}$ preserves mixtures : if $ \alpha = 
( \{\alpha_{i}\}, \{ p_{i} \} )$, we have
\begin{eqnarray*}
\hat{\alpha} (\phi) = \phi (\alpha ) = \sum_{i} p_{i} \phi(\alpha_{i}) = \sum_{i}p_{i}\hat{\alpha_{i}} (\phi ) 
\end{eqnarray*}
implying $ \hat{\alpha} = \sum p_{i} \hat{\alpha_{i}} $.

\noindent
Remarks : (i) We have implicitly used, in the argument above, the fact that $ \mathcal{S} $ is a 
`separated' mixture space which means that, given two states $ \alpha_{1} \neq \alpha_{2} $ in 
$ \mathcal{S} $, there is at least one affine functional $ \phi $ on $ \mathcal{S} $ such that 
$ \phi ( \alpha_{1} ) \neq \phi ( \alpha_{2} )$. This is clear from the definition of states given 
above. [ For given Q and B, the object $ w^{Q}_{.}(B) $ given by $ w^{Q}_{.}(B) (\alpha) = w^{Q}_{\alpha} (B) $ is an affine functional on $ \mathcal{S} $, etc.] \\
(ii) Note that the treatment of states and observables presented above is quite general; nowhere 
in our proceedings did we commit ourselves to any specific type of origin of the probabilistic 
aspect of the phenomena.\\
(iii) The states in classical statistical mechanics ( probability densities in phase space ) already constitute a standard mixture space.\\
(iv) Deterministic theories are a subclass of the theories covered by the present formalism --  
those in which all probabilities are either zero or one. [Recall eq (2).]\\
(v) Note, from eqs (4) and (6A), that an observable Q defines an affine functional $ \phi_{Q} $ on 
$ \mathcal{S} $ ( the expectation value functional ) given by $ \phi_{Q} ( \alpha ) = \alpha (Q). $

\vspace{.2in}
\noindent
B.\textsl{Probabilistic Hamilton-Jacobi Theory ( the Hamilton-Jacobi Fluid [25,28] )}

  Given a system with a finite number of degrees of freedom with a Lagrangian 
$ L (q, \dot{q}, t)$, its state at any time t is given by the collection $ ( q (t), \dot{q}(t))$ 
and its dynamics is given by the Euler-Lagrange equations obtained by applying Hamilton's 
principle to the action $ \int{Ldt}$. An equivalent description is given in phase space in terms 
of $q^{\alpha}$ and $ p_{\alpha} = \partial L /\partial \dot{q}^{\alpha}$; now the state at any 
time is (q(t), p(t)) and dynamics is given by the Hamilton's equations
\begin{eqnarray}
\dot{q}^{\alpha} = \frac{\partial H}{\partial p_{\alpha}} \hspace{.3in} \dot{p}_{\alpha} = 
-\frac{\partial H}{\partial q^{\alpha}} .
\end{eqnarray}

  A third equivalent description is obtained in terms of the Hamilton-Jacobi function S(q,t) 
which, for a fixed time $t_{0}$ and configuration $q_{0}$, is given by 
\begin{eqnarray}
S(q, t) = \int_{q_{0},t_{0}}^{q,t}dt^{'}L( q(t^{'}, \dot{q}(t^{'}),t^{'})
\end{eqnarray} 
where the integration is along the physical trajectory between $(t_{0}, q_{0})$ and (t, q(t)) (assumed, for simplicity, unique). It satisfies the Hamilton-Jacobi equation
\begin{eqnarray}
\frac{\partial S(q, t)}{\partial t} + H (q, \frac{\partial S(q,t)}{\partial q},t) = 0 .
\end{eqnarray}
A solution S(q,t) of eq(11) with the initial condition $ S(q,t_{0}) = S_{0}(q) $, when supplemented 
with the initial condition
\begin{eqnarray}
q(t_{0}) = q_{0},
\end{eqnarray}
can be used to obtain the (unique) dynamical trajectory given by the Hamilton's equations (9) with 
initial condition $ (q(t_{0}),p(t_{0}))=  (q_{0},p_{0}) $ where $ p_{0 \alpha} = 
\frac{\partial S_{0}(q)} {\partial q^{\alpha}} \mid_{q = q_{0}} $. To see this, define the momentum 
field p (q,t) and the velocity field v(q,t) by
\begin{eqnarray}
p_{\alpha}(q,t) & = & \frac{\partial S(q,t) }{\partial q^{\alpha}} \\
v^{\alpha}(q,t) & = & \frac{\partial H(q,p,t)}{\partial p_{\alpha}} \mid _{p = p(q,t)} .
\end{eqnarray}
The differential equation
\begin{eqnarray}
\dot{q}^{\alpha} (t) = v^{\alpha}(q(t),t) 
\end{eqnarray}
with the initial condition (12) gives the unique solution $ q^{\alpha}(t). $ Finally 
\begin{eqnarray}
p_{\alpha}(t) = p_{\alpha}(q(t),t).
\end{eqnarray}

  Note that, in this picture, the state at time t is given by the quantities $ q^{\alpha}(t) $ and 
S(q,t). The corresponding phase space density function is 
\begin{eqnarray}
\tilde{\rho}(q,p,t) = \delta( q - q(t)) \delta \left(p - \frac{\partial S(q,t)}{\partial q} \right) .
\end{eqnarray})
Given a  Hamiltonian H(q,p,t), the change of state with time is given by the differential 
equations (15) and (11). The particle picture ( in situations when the system consists of one or 
more particles) can be recovered from the field S(q,t) through the equations (15) and (16). It 
is instructive to note that, even in classical mechanics, the description of state of a particle 
system can involve a field like S(q,t) in configuration space.

  If, instead of the condition (12), we are initially given a probability distribution $ \rho(q,
t_{0}) = \rho_{0}(q) $, we shall obtain, instead of the functions $ q^{\alpha}(t) $, the 
probability density function $ \rho(q,t) $ which is obtained as a solution of the continuity 
equation
\begin{eqnarray}
\frac{\partial \rho (q,t)}{\partial t} + \frac{\partial}{\partial q^{\alpha}} [v^{\alpha}(q,t)
\rho (q,t) ] = 0
\end{eqnarray}
with the given initial condition. The corresponding phase space density function is 
\begin{eqnarray}
\tilde{\rho}(q,p,t) = \rho (q,t) \delta (p - \frac{\partial S(q,t)}{\partial q} ) .
\end{eqnarray}
We now have the state at time t described by the pair of fields $ \rho(q,t) $ and S(q,t) whose 
change with time is described by eqs (18) and (11). We have  (mathematically) a 
hydrodynamics-like situation. ( Hence the name Hamilton-Jacobi fluid for this system. ) The 
results at any stage in the formalism can be analysed in terms of a particle picture through the equations (15) and (16).
 
\begin{center}
III. UNIFICATION OF DYNAMICS AND PROBABILITY IN CONFIGURATION SPACE; RATIONALE FOR A COMPLEX
HILBERT SPACE
\end{center}

  As stated in the introduction, in the author's opinion, the main unifying concept in 
quantum mechanics is the unification of dynamics and probability. In this section it will be 
shown that when this unification is considered in the context of particle dynamics at the 
configuration space level, the wave mechanical formalism of Schr$\ddot{o}$dinger naturally 
emerges.

\vspace{.2in}
\noindent
A.\textsl{States}

  We are looking for a unification of the kind one has, for example, in classical relativistic
electrodynamics : the electric field $ \vec{E} $ and the magnetic field $ \vec{B} $ are 
unified into the electromagnetic field tensor $ F_{\mu \nu} $. Here both the objects being 
unified are fields; the unified object is also a field having a `higher' mathematical 
structure. Our search for unification of probability and dynamics should begin with a 
formulation of the probalistic version of classical dynamics in which aspects of probability and 
dynamics are both represented by similar mathematical objects. In phase space such a 
unification is already achieved by the phase space probability density function 
$ \tilde{\rho}$ (q,p,t ). We would like to achieve it at the level of configuration space 
(which is a more fundamental level ). Here the choice is obvious : the probalistic version 
of Hamilton - Jacobi theory discussed in section II B; here  probability and dynamics 
are represented by the fields $ \rho $ (q,t ) and S(q,t) respectively. Restricting, for simplicity, to single 
particle systems, we attempt to replace $ \rho (x,t) $ and S (x,t) by a single function 
F(x,t) which presumably represents, in the quantum theory to be evolved, the state of the 
particle at time t ( incorporating the statistical features of its kinematics at time t ).
One should keep in mind the possibility that the function F may belong to a `higher'
category of mathematical objects than that of $ \rho $ and S ( which are real - valued 
functions of their arguments ). Having arrived at the class of mathematical objects to which 
F must belong, we shall attempt a straightforward treatment of states and observables in the 
framework of section IIA.     

  In the emerging quantum kinematics and dynamics, the Planck constant $ \hbar $ is expected to 
appear in various quantities. We expect the function F(x,t) to involve, as parts of its structure,
 two functions $ \tilde{\rho}$(x,t) (this has nothing to do with the phase space density function 
$ \tilde{\rho}$ used earlier ) and $ \tilde{S} $(x,t) such that, in the limit $ \hbar \rightarrow 0$
\begin{eqnarray}
\tilde{\rho}(x,t) \rightarrow \rho (x,t) \hspace{.3in} \tilde{S}(x,t) \rightarrow S(x,t) .
\end{eqnarray}

  Let F = f ($ \tilde{\rho}, \tilde{S} $ ). To determine the function f, we note that, for two non-
interacting particles 1 and 2, the functions $ \rho_{12} (x^{(1)}, x^{(2)}, t) $ and 
$ S (x^{(1)}, x^{(2)}, t) $ for the two-paticle system are related to the functions $ \rho_{i} $ 
and $ S_{i} $ (i = 1,2 ) for the two one-particle systems by 
\begin{eqnarray}  
\rho_{12} (x^{(1)}, x^{(2)}, t) & = & \rho_{1}(x^{(1)},t) \ \rho_{2}(x^{(2)},t) \\
      S_{12}(x^{(1)},x^{(2)},t) & = & S_{1}(x^{(1)},t) +  S_{2}(x^{(2)},t) .
\end{eqnarray}
It appears reasonable to assume that the functions $ \tilde{\rho} $ and $ \tilde{S} $ will also satisfy the conditions (21) and (22 respectively. Now the function 
$ f_{12}(\tilde{\rho}_{12},\tilde{S}_{12}) $ must be related to $f_{1}(\tilde{\rho}_{1},\tilde{S}_{1}) $ and $f_{2}(\tilde{\rho}_{2},\tilde{S}_{2})$ in a definite way. The simplest possibilities are 
\begin{eqnarray}
f_{12}(\tilde{\rho}_{12},\tilde{S}_{12}) = f_{1}(\tilde{\rho}_{1},\tilde{S}_{1}) f_{2}(\tilde{\rho}_{2},\tilde{S}_{2}) 
\end{eqnarray}
and
\begin{eqnarray}
f_{12}(\tilde{\rho}_{12},\tilde{S}_{12}) = f_{1}(\tilde{\rho}_{1},\tilde{S}_{1}) + f_{2}(\tilde{\rho}_{2},\tilde{S}_{2}) .
\end{eqnarray}
In fact, these two possibilities are mutually related: given (23), g = ln f satisfies (24) and 
given (24), h = exp( f) satisfies (23). We assume that (23) holds. Eqs(21--23) give
\begin{eqnarray}
f(\tilde{\rho}, \tilde{S}) =  (\tilde{\rho})^{a} exp[b \tilde{S}]
\end{eqnarray}
where a and b are constants.

  Assuming that $ \tilde{\rho} $ is also a probability ensity, it is reasonable to demand 
that, given F(x,t), the quantity $\tilde{\rho}$(x,t) is uniquely determined.(This amounts to 
requiring that, in a given state, there is a unique probability density for the particle position.)The 
simplest way to achieve this is to have the parameter a real and b imaginary (say, b = i$\lambda$ with $\lambda$ real ).Assuming this, we have
\begin{eqnarray}
F(x,t) = [\tilde{\rho}(x,t)]^{a}exp[i\lambda \tilde{S}(x,t)] .
\end{eqnarray}

  Now, the function S(x,t) is arbitrary upto an additive (real) constant; the same is also 
expected of $\tilde{S}$. It follows that a multiplicative constant phase factor in  
F(x,t) must be inconsequential for the representation of state.
 The objects which uniquely determine the state at time t are the bilocal functions
\begin{eqnarray} 
w(x,x^{'},t) = F(x,t)F^{*}(x^{'},t) = [\tilde{\rho}(x,t)\tilde{\rho}(x^{'},t)]^{a}exp[i\lambda\{\tilde{S}(x,t) - \tilde{S}(x^{'},t) \}] .
\end{eqnarray}
Note that
\begin{eqnarray}
w(x,x,t) = [\tilde{\rho}(x,t)]^{2a} .
\end{eqnarray}

  Recalling the discussion in section IIA, we now impose the requirement that convex 
combinations of quantities of the form (27) must be admissible states. let $ 0 < p < 1$ and
\begin{eqnarray}
w_{12}(x,x^{'},t) = p w_{1}(x,x^{'},t) +(1-p)w_{2}(x,x^{'},t)
\end{eqnarray}
where $w_{1}$ and $w_{2}$ are of the form (27). Putting x = $x^{'}$ in eq(29), we get
\begin{eqnarray}
w_{12}(x,x,t) = p[\tilde{\rho}_{1}(x,t)]^{2a} + (1-p)[\tilde{\rho}_{2}(x,t)]^{2a} .
\end{eqnarray}
Now a convex combination of probability densities is a probabilty density. Eq(30),
therefore, appears to make sense only if 2a = 1. Assuming this, we have, finally, (replacing the symbol F by the conventional $\psi$ ) the Schr$\ddot{o}$dinger type wave function
\begin{eqnarray}
\psi(x,t) = [\tilde{\rho}(x,t)]^{1/2} exp[i\lambda \tilde{S}(x,t)] .
\end{eqnarray}
With $|\psi (x,t)|^{2} = \tilde{\rho}(x,t)$ the probability interpretation of $\psi$ is 
automatic and 
\begin{eqnarray}
\int|\psi(x,t)|^{2}dx = \int \tilde{\rho}(x,t)dx = 1 
\end{eqnarray}
so tht, for each fixed t, the functions $\psi (.,t)$ belong to the Hilbert space 
$\mathcal{H}$ of complex square integrable functions on $R^{3}$.

  The quantities $w(x,x^{'},t)$ (for general states including mixed states ) satisfy the 
conditions
\begin{eqnarray}
w(x,x^{'},t)^{*} = w(x^{'},x,t)  \hspace{.3in} \int w(x,x,t) dx = 1 
\end{eqnarray}
so that they are kernel functions of density operators on $\mathcal{H}$ :
\begin{eqnarray}
w(x,x^{'},t) = <x^{'}|w(t)|x> .
\end{eqnarray}

\noindent
Note. The functions $\psi (.,.)$ are, indeed, mathematical objects belonging to a `higher' 
category than that of $\rho$ and S : they are complex valued functions. In the present setting, the appearance of complex valued functions should not be surprising: fluctuations are
going to be an important part of the physics to emerge and these are most conveniently 
analysed in a complex variable setting.

  An importnt consideration about states is the differentiability requirements on the wave 
functions $\psi$. In classical mechanics, the functions $\rho$(x,t) and S(x,t) are smooth. These 
functions, however, are (supposedly) limits of the functions $\tilde{\rho}$ and $\tilde{S}$ in 
the limit of vanishing Planck constant. The smoothness of $\tilde{\rho}$ and $\tilde{S}$ or of 
$\psi$ must be decided directly on the basis of appropriate physical and mathematical 
consierations  in the quantum mechanical theory we are trying to construct, without any 
reference to the classical theory. Physically, the wave functions are supposed to encode the 
data about preparation of the system in question. Since any laboratory preparation involves some 
error margins, the objects employed in the encoding should be, broadly speaking, reasonably 
smooth so that small changes in the input data imply small change in the wave function ( hence 
in the statistical information provided by it ). We shall, therefore, assume that the physically 
realizable pure states are represented by sufficiently smooth wave functions; we shall denote this subclass of functions in $\mathcal{H}$ by $\mathcal{S}_{1}$.(More precise specification of \sone will appear later.) 
Since linear combinations of smooth square integrble functions are smooth square integrable 
functions, $\mathcal{S}_{1}$ must be a vector subspace of $\mathcal{H}$. For 
analytical work one must include the limits of sequences of functions in $\mathcal{S}_{1}$ and 
consider its completion. This completion must be $\mathcal{H}$ (there being no grounds for 
taking  it to be a proper subset of $\mathcal{H}$); $\mathcal{S}_{1}$ must, therefore, be a 
dense subspace of $\mathcal{H}$. The space of all physical states ( including mixed states 
constructed from pure physical states in $\mathcal{S}_{1}$ ) will be called $\mathcal{S}$.

\noindent
Remarks. (i) Smoothness of $\psi$ has nothing to do with any possible smoothness of allowed 
particle trajectories. Even in classical mechanics, the smoothness of S(x,t) has nothing to do with 
smoothness of particle trajectories. It has, rather, to do with the fact that the Hamilton-Jacobi function is given by the 
integral in eq (10).
In fact, there is a formalism [35] in which the wave function $\psi$ is defined as an 
integral over (continuous) particle trajectories; its smoothness is then almost automatic. \\
(ii)The vector space nature of $\mathcal{S}_{1}$ implies the principle of superposition of 
quantum mechanical (pure) states : given two elements $\psi_{1}$ and $\psi_{2}$ of 
$\mathcal{S}_{1}$ (each of which represents a quantum mechanical pure state upto a constant phase 
factor) the superposition $ \psi = a\psi_{1} + b\psi_{2}$ (with $|a|^{2} + |b|^{2} = 1$) also 
represent a quantum mechanical pure state upto a constant phase factor. \\
(iii) The space $\mathcal{S}_{1}$ is the quantum mechanical analogue of the classical phase 
space of a system. 

\vspace{.2in}
\noindent
B. \textsl{Observables}

  Recalling eq(4) and the remark (v) in section IIA, we expect observables to be objects defining 
affine functionals on the space $\mathcal{S}$ of physical states. Suppressing time argument, a 
general physical state is of the form $ \rho(x,y) = \sum p_{i} \rho_{\psi{i}}(x,y)$  where 
$ \rho_{\psi}(x,y) = \psi(x) \psi^{*}(y)$ and the $\psi$s are in $\mathcal{S}_{1}$. An affine 
functional on these objects is given by a complex bilocal function A(y,x) giving the expectation 
value functional 
\begin{eqnarray}
<A>_{\rho} = \rho(A) = \int A(y,x)\rho(x,y)dxdy = \sum_{i} p_{i} \int \psi_{i}^{*}(y) A(y,x) \psi_{i}(x)dxdy.  
\end{eqnarray}
It is adequate to consider the pure state expectation values
\begin{eqnarray}
<A>_{\psi} = \int \psi^{*}(y)A(y,x) \psi(x) dxdy .
\end{eqnarray}
Keeping in mind the denseness of $\mathcal{S}_{1}$,  reality of $<A>_{\psi}$ for all $\psi$ in 
$\mathcal{S}_{1}$ implies $A(y,x)^{*} = A(x,y)$ so that A(x,y) is the kernel of a self-adjoint 
operator :
\begin{eqnarray*}
A(y,x) = <y|A|x> \hspace{.2in} A = A^{\dagger} \hspace{.2in} (A \psi)(y) = \int A(y,x) \psi (x) dx .
\end{eqnarray*}
\noindent
Note. We did not show in section IIA that every affine functional on the set of states defines an 
observable. In the case at hand, however, we can show directly that the self adjoint operator A above defines 
an observable  in the sense of section IIA. For this, it is enough to show that, given any pure 
state $\psi$, a self ajoint operator defines a probability measure $ w^{A}_{\psi}$ on the real line 
R. Given a self-adjoint operator A and a Borel measurable function f on R, the operator f(A) can be 
defined through the spectral theorem. Now we can define $w^{A}_{\psi}$ by 
\begin{eqnarray*}
w^{A}_{\psi}(B) = (\psi, \chi_{B}(A) \psi) \  for\  every\  Borel\  set\  B.
\end{eqnarray*}
The self adjoint operators corresponding to the expectation value functionals, therefore, qualify 
to be called observables.

  Since the operation of the operator A above needs to be defined only on the wave functions in 
$\mathcal{S}_{1}$, it is allowed to be an unbounded operator; had we defined observables in terms of 
states on all of $\mathcal{H}$, we would end up with bounded self adjoint operators as observables.

\noindent
Note. Existence of the integral in eq(36) for $\psi$ in $\mathcal{S}_{1}$ does not demand that 
the vector $A\psi$ also lie in $\mathcal{S}_{1}$. Since, however, $\mathcal{S}_{1}$ is dense, nothing 
of significance is lost in assuming  that the self adjoint operators representing observables map 
$\mathcal{S}_{1}$ into itself. We shall henceforth assume this; this is to ensure that the 
observables belong to the algebra \aq \  defined below.

  The collection $\mathcal{A}_{Q}$ of linear operators which, along with their adjoints, 
map \sone \ into itself, is an associative *-algebra ( the *-operation being hermitian 
conjugation ). This object is the analogue of the algebra \acl \ of (smooth) comlex valued functions 
on phase space in classical dynamics and will play an important role in our treatment of 
quantum kinematics. Note that, whereas \acl \ is a commutative algebra [with product of functions 
defined as fg(q,p) = f(q,p) g(q,p)], the algebra \aq \ is noncommutative.This is a concrete formulation of Heisenberg's 
insight [5] that kinematics underlying quantum dynamics must be based on a noncommutative algebra of observables.

\noindent
Remarks.(i) There is, as we shall see, some flexibility in the choice of the spaces \sone\  
and \aq. We shall eventually find it convenient to define \sone\  as the largest common dense domain 
which is mapped into itself by the so-called `fundamntal observables' (these are defined 
in section V ) and \aq \ as the *-algebra generated 
by these observables. This has the advantage that \aq \ so defined has a trivial 
center (i.e. all operators commuting with every operator in \aq \ are multiples of the 
identity operator ).   \\
(ii) One can introduce, at this stage, a topology $\tau$ on the space \sone\  such that the pair
$ (\sone, \tau ) $ becomes a topological vector space and the  
operators in \aq \ are continuous operators on this topological vector space. 
Let $ \sone^{*}$ be the topological dual of $ (\sone,\tau$) i. e. the space of continuous linear 
functionals on \sone. We then have (with a little bit more of mathematical finesse ) the 
rigged Hilbert space or Gelfand triple [36--38]
\begin{eqnarray*}
\sone \subset \hs \subset \sone^{*}
\end{eqnarray*}
($\sone^{*}$ is the space to which the generalized eigenvectors of the operators in \aq 
\ belong--for example, the generalized eignfunctions $e^{ikx}$ of the momentum operator
-id/dx; these functions obviously don't belong to \hs.) One can then have a 
mathematically rigorous development [39, 40, 38] of the Dirac bra-ket formalism [41]. We 
shall skip the details.

\vspace{.2in}
\noindent
C. \textsl{ Transition Probabilities; Fundamental Invariances }

  In any scheme of dynamics, geometrical properties of the basic spaces play an important role. 
In classical mechanics, for example, the symplectic structure on a phase space and related 
canonical transformations play very important role. Among the two basic spaces, \sone\  and \aq 
\ introdued above, we consider here the geometry of \sone\ ; that of \aq \ (which will involve 
noncommutative geometry ) will be taken up in section V.

  Apart from the vector space structure (which, as we have seen, implies the principle of 
superposition), the space \sone\  has a scalar product defined on it. The first question we must 
consider is the physical significance of the quantity $ (\phi, \psi)$ for $\phi$ and $\psi$ in 
\sone . Note, in this connection, that the state represented by $\phi$ can also be equivalently 
represented by the projection operator $P_{\phi}$ for $\phi$. Since $P_{\phi}$ is a self-adjoint 
operator belonging to \aq, it is an observable; it tests whether or not the given state is $\phi$. 
The expectation value of $P_{\phi}$ in the state $\psi$ is easily sen to be $ |(\phi, \psi)|^{2}$; 
the natural interpretation of this quantity is the probbility that, given the system in the state 
$\psi$, on measuremnt it is found to be in the state $\phi$ (transition probability from the 
state $\psi$ to $\phi$ which happens to be equal to that from $\phi$ to $\psi$ ). The quantity 
$(\phi,\psi)$( the orthogonal component of $\psi$ along $\phi$ )  is, therefore, given the name 
`transition amplitude from the state $\psi$ to the state $\phi$'.

  Note that, whereas in classical stochastic theory (for example, in the context of Markov 
processes [42--44]) one only talks about transition probabilities, in QM we have transition 
amplitudes as well. This feature is, of course, closely related to the principle of superposition.

  Transformations on states which leave transition probabilities invariant are traditionally 
considered as fundamental invariances of the quantum mechanical formalism. according to Wigner's 
theorem [45--47], an invertible transformation on \sone\  (which, by continuity, can be extended to 
\hs ) mapping a state $\psi$ to $\psi_{\prime}$ such that
\be
|(\phi^{'},\psi^{'})|^{2} = |(\phi,\psi)|^{2}
\ee
can, by appropriate choice of phases in the representation of states (by vectors in \hs) be 
represented by a unitary or an antiunitary transformation ($ \psi^{'} = U\psi $ where U is unitary 
or antiunitary ).

\noindent
Note. If we stick to the convention that only the transformations leaving the fundamental 
geometrical structure (scalar product or, equivalently, transition amplitudes, in the present 
case) invariant are to be called fundamental invariances, then only the unitary transformations 
in  the conclusion of the above theorem qualify to be fundamental invariances. These are the  
analogues  of canonical transformations in classical mechanics.The antiunitary ones (which leave
the transition probabilities but not the transition amplitudes invariant ) also happen to 
be of coniderable importance because, in practice,  transition probabilities are more important objects than the phases of transition amplitudes. We shall see in section V that it is the unitary 
transformations only that qualify as the quantum mechanical canonical transformations defined as 
invariances of the noncommutative symplectic structure on \aq.

\begin{center}
IV. GENERALIZED ALGEBRAIC SYMPLECTIC SRTUCTURES
\end{center}

  In this section we shall construct, employing nonommutative differential geometric techniques, 
a class of mathematical objects which can accommodate, as a special case, the classical 
Hamiltonian systems and also provide the proper setting for a satisfactory treatment of quantum 
symplectics.

  First, a few algebraic preliminaries.

\vspace{.2in}
\noindent
A. \textsl{ Algebras and Derivations}

  By an algebra we shall mean a complex associative algebra with unit element ( usually denoted 
as I ) and a *-operation (involution ). We shall denote algebras by script letters \sca, \ \scb,.
.. and elements of an algebra by capital letters A,B,...The star operation, by definition satisfies the relations 
\be
(AB)^{*} = B^{*} A^{*} \hspace{.2in} (A^{*})^{*} = A \hspace{.2in} I^{*} = I
\ee

  A (*- ) homomorphism of an algebra \sca \ into \scb \ is a linear mapping $ \Phi : \sca \rightarrow \scb $ which preserves products 
( and involutions ) :
\be
\Phi(AB) = \Phi(A) \Phi (B) \hspace{.4in} \Phi (A^{*}) = \Phi(A)^{*} ;
\ee
if it is, moreover, bijective, is called a (*-) isomorphism.

  A derivation of an algebra \sca \ is a linear mapping $ X :\sca \rightarrow \sca $ obeying the 
Leibnitz rule 
\be
X(AB) = X(A)B + A X(B).
\ee
The set Der\sca \ of all derivations is a lie algebra ( with commutator as the Lie bracket ). The 
inner derivations $ D_{A} $ of \sca \ defined by 
\be
D_{A} B = [A,B]
\ee
satisfy the relation
\be
[D_{A},D_{B}] = D_{[A,B]}
\ee
so that the set $ IDer\sca$ of inner derivations of \sca \ is a Lie subalgebra of $Der\sca$.

  An algebra isomorphism $\Phi : \sca \rightarrow \scb $ induces a mapping $ \Phi_{*} : Der\sca 
\rightarrow Der\scb $ given by 
\be
(\Phi_{*}X )(B) = \Phi (X(\Phi^{-1}(B))
\ee
for all $  X \in Der\sca $ and $ B \in \scb. $ We have the relations 
\begin{eqnarray}
(\Psi \circ \Phi)_{*} & = & \Psi_{*} \circ \Phi_{*} \\
        \Phi_{*}[X,Y] & = & [\Phi_{*} X, \Phi_{*}Y] .
\end{eqnarray}

\noindent
B. \textsl{ Generalized Algebraic Differential Forms }

  The noncommutative generalization of differential geometry is based on the observation that
most of the developments relating to differential forms can proceed in purely algebraic terms 
[ starting with the commutative algebra \cim ] : \\
(i) Vector fields can be obtained as derivations  of \cim. \\
(ii) Definition of differential forms of various degrees can be given in algebraic terms (by 
defining their contractions with vector fields ). \\
(iii) Among the two basic operations on differential forms, the exterior product and the 
exterior derivative, the former is already algebraic; the ltter can also be defined 
algebraically [48] :
\begin{eqnarray}
(d\omega)(X_{1},X_{2},...,X_{k+1}) =  \sum_{i=1}^{k+1} (-1)^{i+1}X_{i}\omega(X_{1},X_{2},..,X_{i-1},X_{i+1},..,X_{k+1}) +   \nonumber \\
\sum_{i<j}(-1)^{i+j} \omega ([X_{i},X_{j}],X_{1},X_{2},..,X_{i-1},X_{i+1},..,X_{j-1},X_{j+1},..,X_{k+1})
\end{eqnarray}
where $\omega$ is a differential k-form and the Xs are vector fields.

  In noncommutative geometry (NCG), one replaces the commutative algebra \cim \ by a general 
complex associative algebra ( not necessarily commutative ). The formalism of NCG closest to 
our 
needs is the one developed by Dubois-Violette and others [2--4]. In their work, derivations of 
the basic algebra  play a role analogous to that of vector fields in traditional 
differential geometry.Several developments proceed parallel to the commutative case. In 
particular, $d\omega$  ( for $\omega$ a noncommutative differential k-form) is defined by 
eq(46) 
where Xs are now derivations. The formalism developed in the present section is a generalization 
of that of these authors. The generalization is base on the obsrvation that, in eq (46), the Xs 
appear either singly or as commutators. It follows that one can restrict the allowed derivations 
to a Lie subalgebra \scx \ of Der\sca \ and develop NCG based on the pair $(\sca, \scx)$; we shall call 
such a  pair an algebraic differential system (ADS). Those ADSs in which \sca \ is noncommutative with a trivial centre and $ \scx = IDer\sca $ will be called special. They will play a special 
role in quantum symplectics.   

  In the construction of tensorial objects on an ADS \ax, the algebra \sca \ plays the role of 
the algebra \cim of smooth functions on a manifold M and \scx \ that of \scx(M), the Lie algebra of smooth vector fields on M. There is one point of contrast between the two situations : whereas 
the product fX of a smooth function and a vector field is a vector field, the product AY of an element A of \sca \ and Y of Der\sca \ is  not a derivation of \sca \ ( except when A is in the centre of 
\sca ).

  A covariant tensor T of rank k (=1,2,...)on \ax is a k-linear mapping of $(\scx)^{k}$ into \sca.
The space of such tensors will be denoted as $\mathcal{T}_{k}\ax$. We define $\mathcal{T}_{0}\ax 
\equiv \sca. $ The interior product $i_{X}$ is defined as usual :
\be
(i_{X} T)( X_{1}, X_{2},...,X_{k-1}) = T(X,X_{1},...,X_{k-1})
\ee
for $k\geq 1$ and $i_{X} T = 0 $ for $T\in \mathcal{T}_{0}\ax $. We have,of course, $ i_{X}^{2} 
= 0. $

  The space $ \Omega^{k}\ax$ \ of  differential k-forms on \ax \ is ,for $k\geq 2 $,  the subspace of 
$\mathcal{T}_{k}\ax$ \ consisting of elements $\omega$ satisfying the antisymetry condition
\be
\omega(X_{\sigma(1)},...X_{\sigma(k)}) = \epsilon_{\sigma} \omega(X_{1},...,X_{k})
\ee
where $\epsilon_{\sigma}$ is the parity/signature of the permutation $\sigma$. We have 
$\Omega^{0}\ax
 = \sct_{0}\ax = \sca $ and $\Omega^{1}\ax = \sct_{1}\ax$.In the notation of ref[3], our 
$\Omega^{k}\ax$ \ is the same as $C^{k}(\scx,\sca)$.

  Exterior product of a p-form $\alpha$ and a q-form $\beta$ is defined as usual [48] (with vector fields replaced by derivations in \scx ):
\be
(\alpha \wedge \beta ) (X_{1},..,X_{p+q})  = \frac{1}{p! q!}\sum_{\sigma \in S_{p+q}}\epsilon_{\sigma}\alpha(X_{\sigma(1)},..,X_{\sigma(p)}) \beta(X_{\sigma(p+1)},..,X_{\sigma(p+q)}) .
\ee
We have the associativity property
\be
(\alpha \wedge \beta) \wedge \gamma = \alpha \wedge (\beta \wedge \gamma)
\ee
and the antiderivation property of $i_{X}$ :
\be
i_{X} (\alpha \wedge \beta) = (i_{X} \alpha) \wedge \beta + (-1)^{p} \alpha \wedge (i_{X} \beta)
\ee
but, in general,  not $ \alpha \wedge \beta = (-1)^{pq} \beta \wedge \alpha.$ ( Recall that the 
differential forms have now values in \sca  \ which need not be commutative.)

  The exterior derivative d : $\Omega^{p}\ax \rightarrow \Omega^{p+1}\ax$ is defined , for p = 0 by 
(dA)(X) = X(A) and, for $p\geq 1$, by eq (46) which, for p = 1 gives
\be
(d \omega ) (X,Y) = X(\omega (Y)) - Y(\omega(X))- \omega([X,Y]).
\ee
We have $d^{2}$ = 0 and the usual antiderivation property for d. As usual, we call a 
differential 
form $\alpha$ closed closed if $d\alpha = 0$ and exact if $ \alpha = d\beta$ for some form 
$\beta$.

\vspace{.2in}
\noindent
C. \textsl{ Induced Mappings and Lie Derivatives}

  An isomorphism between two ADSs \ax and \by is a *-isomorphism $\Phi : \sca \rightarrow \scb $ 
such that the induced mapping $\Phi_{*} :\scx \rightarrow \scy $ is a Lie algebra isomorphism. Such 
a mapping induces a mapping $\Phi^{*} : \sct_{k}\by \rightarrow \sct_{k}\ax $ :
\be
(\Phi^{*}T)(X_{1},..,X_{k}) = \Phi^{-1}[ T(\Phi_{*}X_{1},..,\Phi_{*}X_{k})].
\ee
We have
\begin{eqnarray}
        (\Psi \circ \Phi )^{*} & = & \Phi^{*} \circ \Psi^{*} \\
\Phi^{*} (\alpha \wedge \beta) & = & (\Phi^{*} \alpha ) \wedge (\Phi^{*}\beta) \\
           \Phi^{*} (d \alpha) & = & d (\Phi^{*} \alpha).
\end{eqnarray}

  Now, given an ADS \ax, let $\Phi_{t} : \sca \rightarrow \sca $ be a one parameter set of 
transformations (i.e. ADS isomorphisms ) given, for small t, by
\be
\Phi_{t}(A) = A + t g(A) + \texttt{o}(t)
\ee
where g is some mapping of \sca \ into itself. The condition $ \Phi_{t}(AB) = \Phi_{t}(A) 
\Phi_{t}(B) $ gives g(AB) = g(A)B + Ag(B) implying that g(A) = Y(A) for some derivation Y of 
\sca; we call Y the infinitesimal generator of $ \Phi_{t}$. We restrict ourselves to 
transformations whose infinitesimal genrators are in \scx. The induced mappings can now be 
used to define  Lie derivatives of various objects. For $ X \in \scx $ and $ T \in \sct_{k}\ax $ 
we define the Lie derivatives $L_{Y}X $ and $ L_{Y}T $ by 
\begin{eqnarray}
(\Phi_{t})_{*} X & = & X + t L_{Y}X + \texttt{o}(t) \\
 (\Phi_{t})^{*}T & = & T - L_{Y}T + \texttt{o}(t)
\end{eqnarray}
( A minus sign appears in the second equation because $ (\Phi_{t})^{*} $ is essentially 
 $ (\Phi_{t}^{-1})_{*}. $ Straightforward calculations give 
\begin{equation}
L_{Y}X = [Y,X] 
\end{equation}
\begin{eqnarray}
(L_{Y}T) (X_{1},..,X_{k}) = Y[T (X_{1},..X_{k})]    
- \sum_{i=1}^{k} T(X_{1},..,X_{i-1},[Y,X_{i}],
X_{i+1},..,X_{k}).
\end{eqnarray}
The Lie derivative has the usual properties
\begin{eqnarray}
               [L_{X},L_{Y}] & = & L_{[X,Y]} \\
          {[ L_{X}, i_{Y} ]} & = & i_{[X,Y]} \\
                       L_{Y} & = & i_{Y} \circ d + d \circ i_{Y} \\
               L_{Y} \circ d & = & d \circ L_{Y} \\
L_{Y} (\alpha \wedge \beta ) & = & (L_{Y} \alpha ) \wedge \beta + \alpha \wedge (L_{Y} \beta ).
\end{eqnarray}
An object  (whose Lie derivatives are defined ) is said to be invariant if its Lie derivatives  
with respect to all derivations in \scx \ vanish. 

\vspace{.2in}
\noindent
D. \textsl{ Generalized Algbraic Symplectic Structures }

  We shall now consider noncommutative generalization of classical symplectic geometry (a 
quick summary of which appears in the appendix; the reader is advised to have a quick look at it 
before proceeding further). As we shall see, the main devlopments will be parallel to those in the classical case.

  A 2-form $ \omega $ on an ADS \ax \ is a symplectic form if  it is (i) closed and (ii) 
nondegenerate in the sense [3] that, for every $ A \in \sca, $ there exists a unique 
derivaion 
$ Y_{A}$ in \scx \ such that 
\be
i_{Y_{A}} \omega = - dA. 
\ee
Such a form will be taken to define a symplectic structure on \ax 
\ and the triple $ ( \sca, \scx, \omega )$ will be called a generalized algebraic symplectic system (GASS).

  A symplectic mapping from a GASS \axal \ to \bybe \ is a mapping $ \Phi : \sca \rightarrow \scb$ 
such that (i) it is an ADS isomorphism between \ax \ and \by \ and (ii) $ \Phi^{*}\beta = \alpha.$ 
A symplectic mapping from a GASS onto itself will be called a canonical/symplectic 
transformation. The symplectic form and all its exterior powers are invariant under canonical 
transformations. 

  If $ \Phi_{t}$  is a one-parameter family of canonical transformations generated by $X \in \scx $, then 
the condition $ \Phi_{t}^{*} \omega = \omega $ implies $ L_{X} \omega = 0 $ which, along with 
equations (64) and $ d\omega =0 $ gives
\be
d (i_{X} \omega) = 0.
\ee
A derivation X satisfying eq(68) will be called locally Hamiltonian. The subclass of such 
derivations for which $ i_{X} \omega $ is exact will be called ( globally) Hamiltonian. The 
Hamiltonian derivation $ Y_{A} $ corresponding to $ A \in \sca $ is given by eq (67) [see 
eq (A.3)]. 

  Give two locally Hamiltonian derivations X and Y, we have
\begin{eqnarray}
i_{[X,Y]} \omega = (L_{X} \circ i_{Y} - i_{Y} \circ L_{X} )\omega
  = (i_{X} \circ d + d \circ i_{X} ) (i_{Y} \omega) \nonumber \\ 
  =d (i_{X} i_{Y} \omega ) = d[\omega (Y,X)]
\end{eqnarray}
which shows that the commutator of two locally hamiltonian derivations is a Hamiltonian 
derivation.

  The Poisson bracket (PB) of A,B $\in \sca $ denoted as \{ A,B \} is defined as 
\be
\{ A,B \} = \omega (Y_{A},Y_{B}) = Y_{A}(B) = - Y_{B}(A) .
\ee
It has the usual properties of a PB : antisymmetry, bilinearity (obvious ), Leibnitz 
rule :
\be
\{ A,BC \} = Y_{A}(BC) = [Y_{A}(B)]C + B[Y_{A}(C)] = \{ A,B \}C + B \{ A,C \}
\ee
and Jacobi identity :
\be
0 = \frac{1}{2} d\omega (Y_{A},Y_{B},Y_{C}) = \{ A, \{ B,C \} \} + \ cyclic \ terms
\ee
where eqs (46) and (70) have been used. We also have [recalling eq (69)]
\be
i_{[Y_{A},Y_{B}]} \omega = d [\omega (Y_{B},Y_{A})] = - d(\{ A,B \})
\ee
which gives [recalling eq (67)]
\be
[Y_{A},Y_{B}] = Y_{ \{ A,B \} }
\ee
showing that the mapping $ A \mapsto Y_{A} $ is a Lie algebra homomorphism.

  An element $ A \in \sca $ can act, via $ Y_{A},$ as the infinitesimal generator of 
a one-parameter family of canonical transformations. The change in $ B \in \sca $ due to 
such an infinitesimal transformation is [see eq (57)]
\be
\delta B = t Y_{A}(B) = t \{ A,B \}.
\ee
In particular, if $ \delta B = t I $ ( infinitesimal `translation' in B ), we have
\be
\{ A,B \} = I
\ee
 which is the analogue of the classical PB relation $ \{ p,q \}_{cl} = 1. $

\vspace{.2in}
\noindent 
E.\textsl{ Canonical Symplectic Structure on a Special ADS [10] }

  On a special ADS $ (\sca, IDer\sca ) $  we can define a differential 2-form $ \omega
_{c} $ ( to be called the canonical 2-form ) by
\be
\omega_{c}(D_{A},D_{B}) = [A,B].
\ee
It is easily verified that $ \omega_{c} $ is closed. Moreover, for any $ A \in \sca $,
the equation
\be
\omega_{c} (Y_{A},D_{B}) = -(dA)(D_{B}) = - D_{B}(A) = [A,B]
\ee
has the unique solution $ Y_{A} = D_{A}. $ The form $ \omega_{c} $ is easily seen to be 
invariant (i.e. $L_{X} \omega_{c} = 0$ for all $ X \in IDer\sca) $. The symplectic structure defined above will be called the canonical symplectic structure on the special ADS 
$ (\sca,IDer\sca ). $ The PB is now a commutator :
\be
\{ A,B \} = Y_{A}(B) = D_{A}(B) = [A,B].
\ee
The invariant symplectic structure on the algebra $ M_{n}(C) $ of complex $ n \times n $
matrices obtained in ref [2] is a special case of the canonical symplectic structure on 
special 
ADSs described above.

  If, in this special ADS, instead of $ \omega_{c}, $ we take $ \omega = \beta \omega_{c} $ as the symplectic 
form (where $ \beta $ is a nonzero complex number ), we have
\be
Y_{A} = \beta^{-1} D_{A} \hspace{1.2in} \{ A,B \} = \beta^{-1} [A,B].
\ee
We shall make use of such a symplectic structure in the treatment of quantum symplectics 
in the next section.

\vspace{.2in}
\noindent
F. \textsl{ Dynamical Systems in Algebraic Setting; Generalized Algebraic Hamiltonin Systems }

  In the algebraic treatment of dynamical systems [49--51], the basic object for any system S 
is taken to be an algebra \sca. It is often taken to be a $ C^{*}-$ algebra; we shall, however, 
not put that restriction. Obervables of S are self adjoint elements of \sca \ and states are an 
appropriate class of linear functionals on \sca \ ( details of this are not relevant at this stage ).
Dynamics is generally defined by a one-parameter family of *- automorphisms $ \Phi_{t} : \sca
\rightarrow \sca $. Let X be the infinitesimal generator of this family and $ A(t) = \Phi_{t}(A).$
 The resulting time evolution is governed by the equation
\be
\frac{dA(t)}{dt} = XA(t).
\ee

  This description of dynamics is quite general and covers almost all known forms of dynamics 
(with continuous time parameter ).For example, the dynamics of a general deterministic system 
(whose state space is assumed to be a differentiable n-manifold M ) is given by  the equation 
defining the integral curves of a vector field X in M :
\be
\frac{dx^{i}(t)}{dt} =   X^{i}( x(t) ) \hspace{.2in} i = 1,2,..,n
\ee
with some initial conditions $ x^{i}(0) = x^{i}_{0}.$ [Eq (1), for example, is a special case of 
eq (82) with n = 6.] Now, for any smooth function f on M, we have
\begin{eqnarray*}
\frac{d}{dt}[f(x(t)] = \frac{dx^{i}}{dt} \frac{\partial f}{\partial x^{i}} 
= (X^{i}\partial_{i}f) (x(t)) = (Xf)(x(t)).
\end{eqnarray*}
Let x(0) = x and f(x(t)) = F(x,t) = F(t)(x).Note that F(t),for each value of t, is an element of 
the algebra \cim.The preceding equation can now be written as
\be
\frac{dF(t)}{dt}(x) = [XF(t)](x) \Leftrightarrow \frac{dF(t)}{dt} = XF(t)
\ee
which is of the same form as eq (81) with \sca = \cim.

  The most important subclass of deterministic systems is that of Hamiltonian systems in which 
the underlying state space is a symplectic manifold $ (M,\omega ) $ and the vector field X of 
eqs (82, 83) is the Hamiltonian vector field ( see appendix )$ X_{H} $ corresponing to the 
hamiltonian H of the system. Formally, the triple $ (M, \omega, H)$  is called a Hamiltonian 
system.
The equation of motion in such a system takes the form
\be
\frac{d F(t)}{dt} = X_{H}F(t) = \{ H,F(t) \}_{cl}
\ee
where $ \{ , \}_{cl} $ is the classical Poisson bracket (see appendix ).

  For our purposes it is useful to introduce the concept of a generalized algebraic Hamiltonian 
system (GAHS). We define it to be a quadruple $ (\sca, \scx, \omega, H ) $ whee $ (\sca, \scx, 
\omega ) $ is a GASS and H a self-adjoint element of \sca \ ($ H^{*} = H; $ the Hamiltonian ). The 
dynamics of such a system is given by eq (81) with $ X = Y_{H} $ :
\be
\frac{dA(t)}{dt} = Y_{H} A(t) = \{ H, A(t) \}.        
\ee

     A classical Hamiltonian system $ (M, \omega, H ) $ is easily seen to be the GAHS 
$ (\sca , \scx, \omega, H ) $ with \sca = \cim, and \scx \ the family of Hamiltonian vector fields 
on M. We shall describe, in the next section, quantum dynamics in the framework of 
a GAHS.
  
  In section VI we shall need the concept of an isomorphism of GAHSs. Two GAHSs 
$ (\sca, \scx, \omega, 
H ) $ and $ ( \scb, \scy, \zeta, K) $ are isomorphic if there exists a symplectic mapping 
$ \Phi : (\sca, \scx, \omega) \rightarrow (\scb,\scy, \zeta ) $ which preserves the Hamiltonians 
(i. e. $ \Phi^{*} K = H $).

\begin{center}
V. QUANTUM MECHANICS OF A PARTICLE
\end{center}

 Having assembled the necessary apparatus, we shall now  do quantum symplectics by applying some results obtained in the 
previous section to the algebra 
\aq \ constructed in section III. The task of the essential geometrical study of the two spaces \sone 
\ and \aq \ thus completed, we shall proceed to present the promised development of the quantum 
kinematics and dynamics of a particle. Galilean invariance plays an important role in this enterprise.
We shall briefly review the essential developments relating to the projective representations of 
the Galilean group. Its kinematical subgroup, the Euclidean group, plays an essential role in the 
identification of the fundamental observables.The two pictures of quantum dynamics -- the 
Schr$\ddot{o}$dinge picture and the Heisenberg picture -- naturally emerge, the latter being a special 
case of a GAHS.

\vspace{.2in}
\noindent
A.  \textsl{ The Quantum Symplectic Form }

  As mentioned earlier, we shall eventually take \aq \ to be the *-algebra generated by the 
fundamental obsevables ( to be defined later in this section ) ensuring thereby that it has 
a trivial center. Assuming this property, we can now make use of the  (modified )
canonical symplectic structure on a special ADS. We take our GASS to be $ (\aq,\scx_{Q},
\omega_{Q}) $ where $ \scx_{Q} = IDer\aq $ and $ \omega_{Q} = \beta \omega_{c} $ [where 
$ \omega_{c} $ is the canonical symplectic form on the special ADS $ ( \aq, \scx_{Q} ) $ 
and $ \beta $ is a nonzero complex number.] We shall call $\omega_{Q}$ the quantum symplectic 
form . Note that ,the form $ \omega_{c} $ being dimensionless, \oq \ has the dimension of the 
constant $ \beta.$ The quantum Poisson brackets are [see eq (80)]
\be
\{ A,B \}_{Q} = \beta^{-1} [A,B].
\ee
We shall take $ \beta = -i\hbar $.(This is the only place where we shall put the Planck 
constant  `by hand '; its conventionl appearance at various  places will be automatic.) This 
choice gives \oq the dimension of action ( the same as that of the classical symplectic 
form $ dp \wedge dq $ ) and makes the quantum PBs the famous Dirac PBs [7, 41].

  A canonical transformation for this GASS (to be referred to later as a quantum canonical 
transformation ) is a mapping $ \Phi : \aq \rightarrow \aq $ 
which is an isomorphism of the ADS $ ( \aq , \scx_{Q} )$ satifying the condition $ \Phi^{*}\oq 
= \oq $. 
Now
\be
(\Phi^{*} \oq )(X_{1},X_{2}) = \Phi^{-1} [ \oq (\Phi_{*} X_{1}, \Phi_{*}X_{2})].
\ee
We need to consider only inner derivations. For $ X = D_{A} $ eq (43) gives 
\be
\Phi_{*} D_{A} = D_{\Phi(A)}. 
\ee
With $X_{1} = D_{A} $ and $ X_{2} = D_{B} $ eq (87) now gives
\be
\Phi( \beta [A,B] ) = \beta [\Phi(A), \Phi(B)] \ for \ all \ A,B \ in \ \aq.
\ee

  From the invariance property of the canonical symplectic form discussed in section IV E 
we can conclude that all 1-paramter *-automorphisms of \aq \ generated by the inner 
derivations of \aq \ are canonical transformations. [This is consistent with eq (89).]

  We shall next examine as to what extent the Wigner symmetries correspond to quantum 
canonical 
transformations. A Wigner thansformation on states given by $ \psi \rightarrow \psi_{\prime} 
= U \psi $ (with U unitary  or antiunitary ) induces a transformation on \aq \ given by 
\be
(\psi^{'}, A \psi^{'} ) = (\psi, A^{'}\psi ) \Rightarrow A^{'} = U^{-1} A U.
\ee
With $ \Phi(A) = A^{'} $ of eq (90), eq (89) gives
\be
U[\beta [ U^{-1} A U, U^{-1}BU ] ]U^{-1} = \beta [A,B]
\ee
which is satisfied for general $ \beta $ for unitary U but for only real $ \beta $ for 
antiunitary U. Since $ \beta$ is, in fact, imaginary, it follows that only unitary Wigner 
symmetries correspond to genuine quantum  canonical transformations. (This is  a 
familiar situation in quantum mechanics; the antiunitary transformations do not preserve 
the canonical commutation relations. See, for example, [52], p.641. )

  Under an infinitesimal unitary trnsformation $ U = I + i \epsilon G $ where G is a self-
adjoint operator (belonging to \aq ) we have, for the infinitesimal change in an element 
A of \aq
\be
\delta A = -i \epsilon [G,A] = \epsilon \{ T,A \}
\ee
where $ T = -i\beta G = -\hbar G.$ In physical applications where G and T generally 
correspond to observables, it is generlly preferable  to work with the operator T which has 
the `right' dimension. Accordingly, we shall generally write, for an infinitesimal 
transformation   
\be
U = I -i \frac{\epsilon}{\hbar}T.
\ee
  
\vspace{.2in}
\noindent
B.  \textsl{ The Galilean Group }

  (No new result is obtained in this subsection. The contribution of the author is in 
collecting the right material and putting it in proper context; it serves to make the 
presentation reasonably self- contained.) 
 
  An important demand on the formalism being evolved is that it must obey the principle of 
Galilean relativity; this means that the group of ( proper ) Galilean transformations must 
be implemented unitarily. This, as we shall see, has important implications for both the 
kinematics and dynamics of quantum systems.

  The Galilan transformations map the space-time point (x,t) to $ ( x^{'}, t^{'} ) $ given 
by ( in matrix notation )
\be
x^{'} = Rx + vt + a \hspace{.3in} t^{'} = t + b
\ee
where R is an SO(3) matrix, v and a are vectors and b a real number. These transformations 
constitute a connected 10-parameter Lie group G. Since quantum mchanicl wave functions are 
arbitrary upto  constant phases, the unitary representatives U(g) [where  
g = (R,v,a,b ) ] are expected to constitute aray/projective representation of G :
\be
U(g_{1}) U(g_{2}) = \omega (g_{1}, g_{2}) U(g_{1} g_{2} )
\ee
where $ \omega $ is a phase factor of modulus one. In this subsection, we briefly consider 
these projective representations [49, 50] in a general separable Hilbert space. 

  For an 
infinitesimal transformation with parameters $ \epsilon^{\alpha} (\alpha = 1,...,10),$ we 
have $ U(\epsilon) = I + i \epsilon^{\alpha} T_{\alpha}. $ In a genuine group representation 
( i.e. when $\omega = 1$ in eq (95) ) we have 
\be
[ T_{\alpha}, T_{\beta} ] = i f^{\gamma}_{\alpha \beta } T_{\gamma}
\ee
 where $ f^{\gamma}_{\alpha \beta} $ are the structure constants of G. In a projective 
representation,  the preceding equation must be generalized to 
\be
[ T_{\alpha}, T_{\beta} ] = i f^{\gamma}_{\alpha \beta} T_{\gamma} + c_{\alpha \beta} I
\ee
where $ c_{\alpha \beta} $ are complex constants.  

  The operators U(g) are defined modulo 
multiplicative phase factors which means that one has the freedom to redefine the 
generators :
\be
T_{\alpha} \rightarrow T_{\alpha}^{'} = T_{\alpha} + b_{\alpha}I
\ee
where $ b_{\alpha} $ are real parameters.Using this freedom, the gnerators $ J_{k} $ 
(rotations ),$ P_{k} $ (space translations), H (time  translations ) and $ G_{k} $ 
(Galilean boosts ) can be shown to satisfy the commutation relations [51][recall eq (93)]  
\begin{eqnarray}
            [ J_{j}, J_{k} ] & = & i \hbar \epsilon_{jkl} J_{l} \\
           {[P_{j}, P_{k} ]} & = & 0 \\
          {[ J_{j}, P_{k} ]} & = & i\hbar \epsilon_{jkl} P_{l} \\
          {[ G_{j}, G_{k} ]} & = & 0 \\
           {[J_{j}, G_{k} ]} & = & i \hbar \epsilon_{jkl} G_{l} \\
           {[P_{j}, G_{k} ]} & = & i \hbar m \delta_{jk} I \\
              {[ H, P_{k} ]} & = & 0 \\
              {[ H, J_{k} ]} & = & 0 \\
              {[ H, G_{k} ]} & = & i \hbar P_{k} 
\end{eqnarray}
where m is a real prameter. For m = 0, we have  a usual vector representation of G. The 
equations above show that the ray representations of the Galilean group are characterized by 
a real parameter m. In physical applications, this parameter is expected to correspond to a 
Galilean invariant intrinsic property of the system. For a particle, as we shall see below, 
m is the traditional (Newtonian ) mass.

\vspace{.2in}
\noindent
C.  \textsl{ The Fundamental Observables }

  We next consider the fundamental observables ( FOs ) in the quantum theoretic description of a 
particle. The most fundamental observables are those related to measurement of position. 
Recall that $ |\psi (x) |^{2} $ has the interpretation of probabality density for position 
( we have suppressed the time variable ). The mean value $ \int x_{j} |\psi(x) |^{2} dx $ 
of the jth component of position must be interpreted as the expectation value, in the state 
$ \psi $ ,of an observable represented by a self adjoint operator $ X_{j} $ :
\be
( \psi, X_{j} \psi ) = \int x_{j} |\psi (x) |^{2} dx 
\ee
which suggests the following definition of the operator $ X_{j} $ :
\be
 ( X_{j} \psi ) (x) = x_{j} \psi (x) \hspace{.2in} j = 1,2,3.
\ee
  To determine the other FOs we note from eq (93) that observables are in bijective 
correspondence with ( generators of ) infinitesimal one-parameter transformations on the 
states. The FOs must be determined in such a manner that, together, they serve to define 
the most general ` kinematically permissible' change of state at a fixed time. What 
does `kinematically permissible' mean ? For this we must take guidance from the operative 
principle of relativity which, in the present situation, is Galilean relativity. As we have 
seen, a  wave function belongs to  a projective representation of the Galilean group G. For
the quantum mechanical description of a particle,  this representation must be irreducible. 
( This is because, in a reducible representation, the states can be divided into more than 
one Galilean invariant subsets which is not acceptable for an `elementary system' like a 
particle [52].) Now, among the various Galilean transformations, time translations relate to 
dynamics ( change of state with time ) and the boosts also relate to transformations involving
time. The true `kinematical subgroup' of G is the Euclidean 
group whose commutation relations appear in eqs (99--107 ). The remaining FOs must be 
(provisionally; see below) the 
infinitesimal generators $ P_{k} $ and $ J_{k} $ ; we shall call them the momentum and the 
angular momentum operators.

  To determine the operators $ P_{j}, $ we consider the transformaation law of the wave 
functions under infinitesimal space transltions $ x \rightarrow x^{'} = x + \epsilon.$The 
standard choice for the transformation law 
\be
\psi^{'} (x^{'} ) = \psi (x) \Leftrightarrow \psi^{'} (x) = \psi ( x- \epsilon )
\ee
gives
\be
\delta \psi = - \epsilon_{j} \frac{\partial \psi}{\partial x_{j}} = -\frac{i}{\hbar}
\epsilon_{j}P_{j} \psi (x)  
\ee
implying
\be
P_{j} = -i \hbar \frac{\partial}{\partial x_{j}}.
\ee
Eqs (109) and (112) give the canonical commutation relations
\be
[ X_{j}, P_{k} ] = i \hbar \delta_{jk} I.
\ee
It is worth emphasizing that eq (113) (in fact, all equations in this section) must be 
understood as operator equations in the subspac \sone\  of the quantum mechanical Hilbert space 
$ \hs = L^{2} (R^{3}). $

  Now, the orbital angular momentum operators 
\be
L_{j} = \epsilon_{jkl} X_{k} P_{l} 
\ee
 satisfy the commutation relations (99) and (101). Defining $ S_{k} = J_{k} - L_{k} $ 
( operators representing intrinsic angular momentum or spin ), we have
\be
  [ S_{j}, L_{k} ] & = & 0 \\ 
{[ S_{j}, S_{k} ]} & = & i \hbar \epsilon_{jkl} S_{l}.
\ee
We, therefore, conclude that the fundamental observables in the quantum mechanical 
description of a particle are $ X_{j}, P_{j} $ and $ S_{j} $ ( j = 1,2,3 ). For a 
spinless particle we have $ S_{j} = 0 $ and the fundamental observables are $ X_{j} $ 
and $ P_{j} $ only.

  We now assume that the algebra \aq \ is the *-algebra generated by the fundamental 
observables identified above subject to the commutation relations (100), (113), (116) and the 
relations 
\be
[X_{j}, S_{k} ] = 0 = [ P_{j}, S_{k} ]. 
\ee
Any operator in \aq \ commuting with the generators must be a multiple of the identity; the 
algebra \aq, therefore, has a trivial center.         
       
\vspace{.2in}
\noindent
D. \textsl{ Quantum Dynamics }

  Dynamics involves the change of states/observables with time. The developments in the 
subsection B involving the time translation generator H ( the Hamiltonian operator ) are 
relevant for dynamics. A close look at eq (90) shows that one can apply a transformation
to states leaving operators unchanged or apply it to operators leaving states unchanged;
the two options are equivalent in a certain sense which the equation makes clear. (A 
unitary transormation  applied to both states and operators is essentially 
a change of basis and can have no physical implications. ) There are, accordingly, two 
standard descriptions of dynamics in QM : one ( the Schr$\ddot{o}$dinger picture ) includes 
time dependence in states and the fundamental operators $ X_{j}, P_{j}, S_{j}$ ( and, 
therefore, any function of these not involving any explicit time dependence ) remain time 
independent; the second ( the Heisenberg picture ) puts time dependence in operators and 
the state vectors remain time independent. The two descriptions are related through eq (90) 
where U is now the time evolution operator obtained below [ see eq (119) ].

  In the Schr$\ddot{o}$dinger picture, we have, under an infinitesimal time translation 
$ t \rightarrow t^{'} = t + \delta t $ [recall eq (93)]
\begin{eqnarray*}
\delta \psi (x,t) = \frac{\partial \psi }{\partial t} \delta t
=  - i \frac{\delta t }{\hbar} H \psi(x,t)
\end{eqnarray*}
which gives the Schr$\ddot{o}$dinger equation ( in the general form )
\be
i \hbar \frac{\partial \psi }{\partial t} = H \psi.
\ee
When H is independent of time, we can write [ notation : $ \psi(x,t) = <x|\Psi (t)>$ ]
\be
\Psi (t) = U(t,t_{0}) \Psi (t_{0})  \hspace{.6in} U(t, t_{0}) = e^{-iH(t-t_{0})/\hbar} 
\ee

  In the Heisenberg picture, we have under an infinitesimal  time translation 
\be
\delta A(t) = -i \frac{\delta t}{\hbar} [H, A(t)] 
\ee
which gives the Heisenberg equation of motion
\be
 \frac{dA(t)}{dt} = (-i\hbar)^{-1} [ H, A(t)] = \{ H, A(t) \}_{Q}.
\ee
Note that eq (121) is of the form of eq (85); we have here the GAHS 
$ (\aq,\scx_{Q},\oq, H ). $

\vspace{.2in}
\noindent 
E.  \textsl{ The Hamiltonin Operator }

  The Schr$\ddot{o}$dinger equation (118) or the Heisenberg equation (121 ) give a 
concrete description of dynamics only when the expression for the Hamiltonian operator H 
in terms of the fundamental operators is given. We first consider the case of a free 
spinless particle.A free particle must be seen as a free particle by all Galilean observers;
its dynamics, therefore, has full-fledged Galilean invariance and all the commutation relations 
(99--107) must be operative. According to eq (104) a partcle is characterized by a Galilean 
invariant parameter m which, as we shall see presently, is the traditional mass parameter. 
The fundamental observabls are $ X_{j}$ and $ P_{j}$. To determine H, we shall use the 
commutators involving H [eqs (105--107)]. Eq (107) is useful only if $ G_{k}s$ are explicitly 
known; we shall, therefore determine $ G_{k} $ first.

  Under an infinitesimal Galilean transformation, we have
\begin{eqnarray*}
\delta X_{j} = \delta v_{k} \{ G_{k}, X_{j} \}  = \frac{ i}{\hbar} \delta v_{k} [G_{k}, X_{j} ].
\end{eqnarray*}
This, with $ \delta X_{j} = (\delta v_{j}) t I$ [ hint: $ \delta (\psi, X_{j} \psi) = 
( \psi, \delta X_{j} \psi ) = \delta v_{j} t, $  etc. ]  gives
\be
[ G_{k}, X_{j} ] = -i\hbar \delta_{jk} t I.
\ee
This gives
\be
G_{k} = P_{k} t + G_{k}^{'} \ with \ [ G_{k}^{'}, X_{j} ] = 0.
\ee
Eqs (123) and (104) give 
\be 
G_{k} = P_{ k} t - m X_{k} + G_{k}^{''}
\ee
where the last term commutes with both$ X_{j} $ and $ P_{k} $ and, therefore, must be a multiple 
of the identity operator. Eq (107) now gives
\be
[H, X_{k} ] = - \frac{i \hbar}{m} P_{k} \ee
which implies
\be H = \frac{|\vec{P}|^{2}}{2m} + H^{'} \ with \ [ H^{'}, X_{k} ] = 0. \ee
Eqs (126) and (105) imply  $ [ H^{'}, P_{k} ] = 0; H^{'}, $ therefore, must be a multiple of 
the identity operator. This term, being inconsequential for dynamics, can be dropped giving 
the free particle Hamiltonian
\be H_{0} = \frac{ |\vec{P}|^{2}}{2m} \ee
which has the same form as the free particle Hamiltonian in classical mechanics. Notice that m 
is, inded, the mass parameter. 

  To describe the dynamics of a particle under some given forces, we generally employ, for 
convenience, a definite reference frame. For example, the expression for a central force takes 
a simple form in a frame in which the center of force is at the origin. With the choice of 
frames so restricted, one cannot invoke full Galilean invariance. In particular, eq (105) is 
no longer operative and we are back to eq (126). With $[ H^{'}, X_{k} ] = 0, $  we can write 
$ H^{'} = V ( X) $ giving finally
\be H = \frac{|\vec{P}|^{2}}{2m} + V( X ). \ee
From this point on, the development of QM along the traditional lines can proceed.

\begin{center}
VI. QUANTUM-CLASSICAL CORRESPONDENCE
\end{center}
 
  Quantum-classical correspondence has several aspects some of which continue to be 
investigated. A comprehensive work which reports on some detailed features of quantum-
classical correspondence employing some techniques of noncommutative geometry is ref [53]
which contains detailed references. As stated in the introduction, we shall be concerned 
with showing how, in the $ \hbar \rightarrow 0 $ limit, one recovers classical mechanics in 
some of its aspects (especially its Hamiltonian structure ) from QM .

  First we recall a slightly refined version of the well-known argument [25, 28] leading 
from the real and imaginary parts of the Schr$\ddot{o}$dinger equation (118) with H of 
eq (128) (after the substitution $ \psi = \sqrt{\rho}exp[iS/\hbar] $), in the limit 
$ \hbar \rightarrow 0, $ the equations (18) and (11) of the probabilistic version of 
Hamilton-Jacobi theory.The
refinement consists in noting that, in the above- mentioned substitution, the quantities 
$ \rho $ and S will have, in general some $ \hbar $ dependence. Accordingly, we make the 
substitution (31) (with $ \lambda = 1/\hbar$ ) in the Schr$\ddot{o}$dinger equation and take 
its real and imaginary parts; we get the two equations 
\be \frac{\partial \tilde{\rho}}{\partial t} +  \bigtriangledown.( \tilde{\rho} 
\tilde{v}) = 0 \ where \ \tilde{v} = m^{-1} \bigtriangledown \tilde{S} \\
\frac{\partial \tilde{S}}{\partial t} + \frac{(\bigtriangledown \tilde{S})^{2}}{2m} +
V - \frac{\hbar^{2}}{2m}\frac{\bigtriangleup (\sqrt{\tilde{\rho}})}{\sqrt{\tilde{\rho}}} = 0
\ee
where $ \bigtriangleup $ is the Laplace operator. When, in the limit $ \hbar \rightarrow 0, $ 
the functions $ \tilde{\rho}$ and $ \tilde{S} $ 
have well defined limits ( say, the functions $ \rho$ and S ), eqs (129) and (130) give, in 
this limit, eqs (18) and (11) of the Hamilton- Jacobi theory.

  Our main concern, however, will be to recover classical symplectics from quantum 
symplectics in the limit of vanishing Planck constant. Our strategy will be to start with 
the quantum GAHS treated in the previous section, transform it to an isomorphic GAHS 
involving phase space functions and $ \star $-products (Weyl-Wigner-Moyal formalism [54--56]) and 
show that the subsystem of this latter GAHS which go to  smooth functions in the $ \hbar 
\rightarrow 0 $ limit produce the classical GAHS in the limit. For simplicity (and 
continuity with the previous section ), we restrict 
ourselves to the case of a particle although the results obtained admit trivial 
generalization to systems with phase space $ R^{2n}$.

  Recall that, in the case at hand, we have $ \hs = L^{2}(R^{3}) $, the fundamental 
observables are $ X_{j},P_{j} $, which generate the *-algebra \aq. The space \sone\  of 
physically admissible wave functions consists of the largest common dense domain which 
is mapped into itself by the fundamental observables ( and, therefore, by the elements 
of \aq ); it  contains,  as a dense subspace, the space $ \Omega = C_{0}^{\infty}(R^{3})$ 
of infinitely differentiable functions with compact support. We have the quantum GAHS 
$(\aq,\scx_{Q}, \oq, H)$ with H of eq (128).

  Any $ \psi \in \sone\  $ can be approximated as closely as we wish by an element of 
$ \Omega. $ For A $ \in \aq $ and $ \psi \in \Omega $ we have
\be (A \psi)(x) = \int K_{A}(x,y) \psi(y) dy. \ee
To the operator A corresponds the Wigner function $ A_{W} $ on the phase space $ R^{6} $
given by
\be  A_{W}(x,p) = \int K_{A}(x + \frac{b}{2}, x - \frac{b}{2} ) e^{-ip.b/\hbar}db. \ee
In terms of the Fourier transform $\tilde{A}_{W}$ of $ A_{W} $ defined by 
\be A_{W} (x,p) = \int \tilde{A}_{W}( r, s ) e^{i(r.p + s.x)}dr ds \ee
the operator A can be written as
\be A (X,P ) = \int \tilde{A}_{W}( r, s )e^{i(r.P + s.X)} dr ds. \ee
There is a bijective correspondence $ A \leftrightarrow A_{W}. $

  Introducing, in $ R^{6}, $ the notations $ \xi $ = ( x,p) and $ \sigma ( \xi, \xi^{'} )
 = p.x^{'} - x.p^{'} $ ( symplectic form in $ R^{6} $ ), we have, for A,B $ \in \aq $ 
\begin{equation}
 (AB)_{W}(\xi) = (2\pi)^{-6} \int exp [ -i \sigma (\xi - \eta, \tau)] A_{W} ( \eta + 
\frac{\hbar \tau}{4} ) B_{W} (\eta - \frac{\hbar \tau }{4} ) \\
\equiv (A_{W} \star B_{W}) ( \xi). 
\end{equation}
The product $ \star $ of eq (135) is the $\star$- product of Bayen et al [57] and the twisted 
product of Liu [58].

  Simple reasoning based on eqs (131),(133) and (134) shows [58] that the functions 
$ A_{W}(\xi) $ belong   to the spce $ O_{M}(R^{6})$ (the space of infinitely differentiable 
functions  which, along with their derivatives, when multiplied by any Schwartz function, 
give functions bounded all over $ R^{6} $ ). These functions are known [58] to form a 
complex noncommutative algebra with the star product  as product. Calling 
this algebra (i.e. the algebra of Wignr functions of operators in \aq \ with the star 
product as product ) $ \mathcal{A}_{W}, $
 we have a GAHS $ (\mathcal{A}_{W}, \mathcal{X}_{W}, \omega_{W}, H_{W} ) $ where 
the first two entries represnt the special ADS based on $ \mathcal{A}_{W} $ and 
$\omega_{W} = -i \hbar \omega_{c}. $ This GAHS is isomorphic to the initial quantum GAHS.
( This is because the correspondence $ A \leftrightarrow A_{W} $ is a *-algebra 
isomorphism; the rest is automatic.) Under this isomorphism of GAHSs the quantum mechanical 
PB (86) is mapped to the Moyal bracket [56]
\be \{ A_{W}, B_{W} \}_{M} \equiv (-i\hbar)^{-1} ( A_{W} \star B_{W} - B_{W} \star A_{W} ).
\ee

  For  functions f,g in $\mathcal{A}_{W} $ having no $ \hbar- $ dependence, we have, from 
eq (135)
\be f \star g = fg - (i\hbar/2) \{ f, g \}_{cl} + O ( \hbar^{2} ). \ee
The functions $ A_{W} (\xi) $ will have, in general, some $ \hbar $ dependence and the 
$ \hbar \rightarrow 0 $ limit may be singular for some of them [59]. We denote by 
$(\mathcal{A}_{W})_{reg}$ the subclass of functions in $ \mathcal{A}_{W} $ whose $ \hbar 
\rightarrow 0 $ limits exist and are smooth (i. e. $ C^{\infty} $ ) functions; it is 
easily seen to be a subalgebra of $ \mathcal{A}_{W}. $ Now, if $ A_{W} \rightarrow A_{cl} $
 and $ B_{W} \rightarrow B_{cl} $  as $ \hbar \rightarrow 0 $ then $ A_{W} \star B_{W} 
\rightarrow A_{cl} B_{cl} $; the subalgebra $(\mathcal{A}_{W})$, therefore, goes over, 
in the $ \hbar \rightarrow 0 $ limit , to the commutative algebra $ C^{\infty}(R^{6}) $
with pointwise product as multiplication. The Moyal bracket of eq (136) goes over to the 
classical PB $\{ A_{cl}, B_{cl} \}_{cl}. $ Assuming that $ H_{W} \in 
(\mathcal{A}_{W})_{reg}, $ the subsystem $ (\mathcal{A}_{W}, \mathcal{X}_{W}, \omega_{W},
H_{W} )_{reg}$ goes over to the classicl GAHS $ (\mathcal{A}_{cl}, \mathcal{X}_{cl}, 
\omega_{cl}, H_{cl})  $ where  $ \mathcal{A}_{cl} = C^{\infty}(R^{6}).$

\begin{center}
VII. CONCLUDING REMARKS
\end{center}

\noindent
1. The somewhat trivial looking generlization of admitting Lie subalgebras of Der\sca \ in the
development NCG along the lines of ref [2--4] was quite crucial in evolving a formlism in 
which both classical and quantum symplectics could be described as special cases of a single 
mathematical object :GAHS and the quantum to classical transition could be seen transparently.

\noindent
2. After having arrived at the conclusion in section III that Schr$\ddot{o}$dinger wave 
functions are the appropriate objects for a unified description of probability and dynamics,
one could have taken a different route than the one followed here and go to, for example path 
integrals [35]. However, evolving an autonomous formalism for QM in the path integral
framework is a bird of different feather than obtaining a path integral representation of 
 the general solution of the Schr$\ddot{o}$dinger equation ( which is already quite 
challenging ). The challenge of the former bird is ,in the author's opinion, worth accepting 
and it promises 
to be quite rewarding.

\vspace{.3in}
\begin{center} APPENDIX: SYMPLECTIC MANIFOLDS  \end{center}

  We shall follow the notational conventions of Woodhouse [60]. Other useful references are 
[61--63].

  A symplectic manifold is a pair (M, $\omega$) where M is a smooth manifold ( we shall be 
concerned with finite dimensional manifolds only ) and $\omega$ is a differential 2-form 
(defined everywhere on M ) which is (i) closed (i.e. $ d\omega = 0 $ ) and (ii) nondegenerate
 in the sense that, at each point $ u \in M, $ the mapping $ T_{u}(M) \rightarrow T^{*}_{u}(M)$
 given by $ X(u) \mapsto ( i_{X} \omega)(u)$ (or $ X^{a} \mapsto X^{b}\omega_{ba} $ ) where X 
is a smooth vector field, is an isomorphism ( of vector spaces ). The second reqirement implies 
that the matrix $ (\omega_{ij}) $ must be nonsingular. The dimension of M must clearly be even, 
say, 2n.

  Locally, the symplectic form $ \omega $ can be expressed in terms of canonical coordinates
 $ ( q^{i}, p_{i} ) $ in the form 
\begin{eqnarray*}
\omega = \sum_{i = 1}^{n} dp_{i} \wedge dq^{i}. \hspace{3in} (A1)
\end{eqnarray*}

  Given two symplectic manifolds $ (M,\omega) $ and $ ( M^{'}, \omega^{'} ), $ a  mapping 
$ \Phi : M \rightarrow M^{'} $ is called symplectic if it is a diffeomorphism and preserves 
the symplectic form (i.e. $ \Phi^{*}\omega^{'} = \omega $ );if the two symplectic manifolds 
are the same, then $ \Phi $ is called a canonical transformation.

  If $ \Phi_{t} $ is a one- parameter family of canonical transformations generated by a smooth
 vector field X, then the condition $ \Phi_{t}^{*} \omega = \omega $ implies $ L_{X} \omega = 0 $
 which, with $ L_{X} = i_{X} \circ d + d \circ i_{X} $ and $ d \omega = 0 $ implies
\begin{eqnarray*}
d(i_{X} \omega ) = 0. \hspace{3.6in} (A2)
\end{eqnarray*} 
A vctor field satisfying the condition (A2) is calld locally Hamiltonian. The subclass of such 
vector fields for which $ i_{X} \omega $ is exact are called (globally ) Hamiltonian. Writing 
this exact form as - df, we have a bijective correspondence between smooth functions  
( arbitrary upto aditive constants ) and Hamiltonian vector fields given by ( denoting the 
Hamiltonian vector field corresponding to the smooth function f by $ X_{f} $ )
\begin{eqnarray*}
i_{X_{f}} \omega = - df. \hspace{3.5in} (A3)
\end{eqnarray*} 

  The Poisson bracket (PB) of two smooth functions f and g on M is defined as
\begin{eqnarray*}
\{ f, g \}_{cl} = \omega ( X_{f}, X_{g} ) =  X_{f} (g) = - X_{g} (f). \hspace{.8in} (A4)
\end{eqnarray*}
In local coordinates it is given by
\begin{eqnarray*}
\{ f, g \}_{cl} = \sum_{i = 1}^{n} \left( \frac{\partial f}{\partial p_{i}} \frac{\partial g}
{\partial q^{i}} - f \leftrightarrow g \right) \hspace{1.6in} (A5)
\end{eqnarray*}
It differs by an overall sign from the definition generlly given in classical mechanics 
textbooks. The adopted convention has the virtue that, with it, the mapping $ f \mapsto  
X_{f} $ from smooth functions into the Hamiltonian vector fields is a Lie algebra 
homomorphism :
\begin{eqnarray*}
[ X_{f}, X_{g} ] = X_{ \{f,g \}_{cl}}. \hspace{3.4in} (A6)
\end{eqnarray*}  

\begin{center}
ACKNOWLEDGEMENTS
\end{center}
 
  The author would like to thank Michel Dubois-Violette for his constructive comments on the 
article in ref [10].

\begin{center}
REFERENCES
\end{center} 

\noindent
\begin{enumerate}
\item A.Connes, ``Noncommutative Geometry'', Academic Press, New York, 1994.
\item M. Dubois-Violette, R. Kerner and J. Madore, \textsl{ J. Math. Phys.} \textbf{ 31 }
 (1994), 316.  
\item M.Dubois-Violette, \textsl{ Noncommutative Differential Geometry, Quantum Mechanics and 
Gauge Theory } in Lecture Notes in Physics, vol 375, Spinger Verlag, 1991.
\item M.Dubois-Violette, \textsl{ Some Aspects of Noncommutative Differential Geometry },
q-alg/9511027.
\item W.Heisenberg, \textsl{ Zs. f. Phys. } \textbf{ 33} (1925), 879.
\item M.Born and P. Jordan, \textsl{ Zs. f. Phys. } \textbf{ 34} (1925), 858.
\item P.A.M.Dirac, \textsl{ Proc. Roy. Soc. } \textbf{ A 109} (1926), 642.
\item M.Born, W.Heisenberg and P.Jordan, \textsl{ Zs. f. Phys.} \textbf{ 35 } (1926), 557.
\item A.Dimakis and F.M$\ddot{u}$ller-Hoissen, \textsl{ J. Phys. A: Math. Gen.} \textbf{ 25} (1992),
 5625.
\item Tulsi Dass \textsl{ Noncommutative Geometry and Unified Formalism for Classical and Quantum 
Mechanics } Indian Institute of Tchnology kanpur preprint (1993).
\item L. de Broglie, \textsl{ Comptes Rendus } \textbf{ 177} (1923), 507.
\item E. Schr$\ddot{o}$dinger, \textsl{ Annalen der Physik } \textbf{ 79} (1926), 361.
\item G.W.Mackey, ``Mathematical Foundations of Quantum Mechanics'', benjamin-Cummings, 
Reading,Mass., 1963.
\item J.M.Jauch, `` Foundations of Quantum Mechanics '', Addison Wesley, Reading, Mass., 1968.
\item V.S.Varadarajan, ``Geometry of Quantum Theory'' 2nd ed.,Springer Verlag, New York, 1985.
\item I.E.Segal \textsl{ Ann. of Math.(2)}\textbf{ 48}(1947), 930.
\item I.E.Segal, ``Mathematical Problems of Relativistic Physics'',American  Mathematical Society, 1963.
\item R.Haag, ``Local Quantum Physics'', Springer-Verlag, 1992.
\item G.E.Emch, ``Algebraic Methods in Statistical Mechanics and Quantum Field Theory'', Wiley, 
New York, 1972.
\item O.Bratteli and D.W. Robinson,``Operator Algebras and Quantum Statistical Mechanics I. 
$ C^{*}- $ and $ W^{*}- $ algebras, Symmetry Groups, Decomposition of States'', Springer, New York,
1979.
\item N.N.Bogolubov, A.A.Logunov, A.I.Oksak and I.T.Todorov, ``General Principles of Quantum 
Field Theory'',Kluwer, Dordrecht, 1990.
\item E.Nelson, \textsl{ Phys. Rev.} \textbf{ 150} (1966), 1079.
\item E.Nelson, ``Dynamicl theories of Brownian Motion'', Princeton University press, 1967.
\item E.Nelson, ``Quantum Fluctuations'', Princeton University Press, 1985.
\item F.Guerra, \textsl{ Phys. Rep. } \textbf{ 77} (1981), 263.
\item A.S.Holevo, ``Probabilistic and Statistical Aspects of Quantum Theory'',North Holland,
Amsterdam, 1982.
\item H.Araki, ``Mathematical Theory of Quantum Fields'', Oxford university Press, 1999.
\item Ph. Blanchard, Ph.Combe and W.Zheng, ``Mathematical and Physical Aspects of Stochastic 
Mechanics'', Lecture Notes in Physics, vol 281, Springer, 1987.
\item Tulsi Dass, \textsl{ Noncommutative Hamiltonian Systems and Quantum-Classical 
Correspondence} in ``Quantum Implications'' ed by C.M.Bhandari, University of Allahabad, 2002.
\item Tulsi Dass and Y.N.Joglekar, \textsl{ Annals of Physics} \textbf{ 287} (2001), 191.
\item R.M.F.Houtappel, H.Van Dam and E.P.Wigner, \textsl{ Rev. Mod. Phys.} \textbf{ 37} (1965), 595.
\item J.-M.Souriau, ``Structure of Dynamical Systems, A Symplectic view Of Physics'', 
Birkh$\ddot{a}$user, Boston,1997. 
\item V.Guillemin and S.Sternberg, ``Symplectic Techniques in Physics'', Cambridge University 
Press, 1984.
\item A.Peres, ``Quantum Theory: Concepts and Methods'', Kluwer,Dordrecht, 1993.
\item R.P.Feynman, \textsl{ Rev. Mod. Phys.} \textbf{ 20} (1948), 367.
\item I.M.Gelfand and G.E.Shilov, ``Generalized Functions'', vol. II, Academic Press, 1967.
\item I.M.Gelfand and N.J.Vilenkin, ``Generalized Functions'', vol. IV, Academic Press, 1964.
\item A. B$\ddot{o}$hm, ``The Rigged Hilbert Space and Quantum Mechanics'', Lecture Notes in Physics, vol. 78, Springer, berlin, 1978.
\item J.E.Robers, \textsl{ Jour. Math. Phys.} \textbf{ 7} (1966), 1097.
\item J.P.Antoine, \textsl{ Jour. Math. Phys.} \textbf{ 10} (1969), 53, 2276.
\item P.A.M.Dirac, ``The Principles of Quntum Mechanics'', 4th ed., Clarendon Press, Oxford, 1958.
\item C.W.Gardiner, ``Handbook of Stochastic Methods for Physics, Chemistry and the Natural 
Sciences'', Springer, Berlin, 1983.
\item L.Arnold, ``Stochastic Differential Equations: Theory and Applications'' John Wiley and Sons,
New York, 1971.
\item K. Yosida, ``Functional Analysis'', 5th ed., Narosa Publishing House, New Delhi, 1979.
\item E.P.Wigner, ``Group Theory and its Application to the Quantum Mechanics of Atomic Spectra'',
Academic Press, New York, 1959.
\item A. Messiah, ``Quantum Mechanics'', vol II, North Holland, Amsterdam, 1970.
\item V.Bargmann,\textsl{ J. Math. Phys.} \textbf{ 5} (1964), 862.
\item Y.Matsushima, ``Differentiable Manifolds'', Marcel Dekker, New York, 1972.
\item V.Bargmann, \textsl{ Annals of Math.} \textbf{ 59} (1954), 1.
\item M.Hamermesh, ``Group Theory and its Application to Physical Problems'', Addison-Wesley, 
Reading, Mass., 1962.
\item A. Katz, ``Classical Mechanic, Quantum Mechanics, Field Theory'', Academic Press, New York, 1965.
\item E.P.Wigner, \textsl{Ann. Math. } \textbf{40}, No 1 (1939). 
\item J.Bellissard and M.Vittot, {\textsl Ann. Inst. Henri Poincar$\acute{e}$ }{\textbf 52}(1990),175.
\item H.Weyl, ``Theory of Groups and Quantum Mechanics'', Dover, New York, 1949.
\item E.P.Wigner, \textsl{ Phys.Rev.} \textbf{ 40} (1932), 749.
\item J.E.Moyal,\textsl{ Proc. Camb. Phil. Soc.} \textbf{ 45} (1949), 99.
\item F.Bayen et al,\textsl{ Annals of Phys.} \textbf{ 110} (1978), 61, 111.
\item K.C.Liu, \textsl{ J. Math. Phys.} \textbf{ 17} (1976), 859.
\item M.Berry, \textsl{ Some Quantum- Classical Asymptotics} in ``Chaos and Quantum Physics'', 
Les Houches, sessionLII,1989, ed. by M.-J.Giannoni, A.Voros and J.Zunn-Justin, Elsevier Science 
Publishers,B.V., 1991.           
\item N.Woodhouse, ``Geometric Quantization'', Clarendon Press, Oxford, 1980.
\item V.I.Arnold, ``Mathematical Methods of Classical Mechanics'', Springer, New York, 1978.
\item Tulsi Dass, ``Symmetries, Gauge Fields, Strings and Fundamental Interactions'' vol. I :
``Mathematical Techniques in Gauge and String Theories'', Wiley Eastern Limited, New Delhi, 1993.
\item Y.Choquet Bruhat, C.Dewitt-Morette and M.Dillard-Bleick, ``Analysis, Manifolds and 
Physics'',North Holland, Amsterdam, 1977. 
    
\end{enumerate}
\end{document}